\documentclass[twocolumn]{aastex61}

\received{--- --, 2017}
\revised{--- --, 2017}
\accepted{\today}

\usepackage[nolist]{acronym}
\usepackage{array}
\usepackage{dashrule}
\usepackage{amsmath}
\usepackage{longtable}
\usepackage{hyperref,graphicx}
\hypersetup{colorlinks=true, citecolor=blue, linkcolor= magenta}
\usepackage{colortbl}

\begin{acronym}[TDMA]
 \acro{LMXB}{Low mass X-ray binary}
 \acrodefplural{LMXB}{Low mass X-ray binaries}
 \acro{LMXBT}{low mass X-ray binary transient}
 \acro{AMXP}{Accreting millisecond X-ray pulsar}
 \acro{RXTE/PCA}{the Rossi X-ray timing explorer/proportional counter array}
 \acro{RXTE}{the Rossi X-ray timing explorer}
 \acro{MAXI}{the monitor of all sky X-ray image}
 \acro{HMXB}{high mass X-ray binary}
 \acrodefplural{HMXB}{High mass X-ray binaries}
 \acro{Swift/XRT}{the Swift gamma-ray burst mission/X-ray telescope}
 \acro{XSPEC}{an X-ray spectral fitting package}
 \acro{FRED}{fast-rise-exponential-decay}
 \acro{DIM}{the disk instability model}
 \acro{LIS}{low-intensity-state}
 \acro{SXT}{soft X-ray transient}
 \acro{HS}{high/soft state}
 \acro{LH}{low/hard state}
 \acro{ASM}{all-sky Monitor}
 \acro{PCU}{proportional counter unit}
 \acro{ISS}{international space station}
 \acro{MSP}{millisecond pulsar}
 \acro{NS}{neutron star}
\end{acronym}

\shorttitle{Partial accretion in the propeller stage of Aql~X--1}
\shortauthors{g\"{u}ng\"{o}r et al.}

\begin{document}

\title{Partial accretion in the propeller stage of low mass X-ray binary Aql~X--1}

\correspondingauthor{C.\ G\"{U}NG\"{O}R}
\email{gungorcan@itu.edu.tr}

\author{C.\ G\"{u}ng\"{o}r}
\affiliation{\.{I}stanbul Technical University, Faculty  of Science  and  Letters,  Physics Engineering  Department,
  34469,  \.{I}stanbul, Turkey}
\affiliation{Sabanc{\i} University,
  Faculty of Engineering and Natural Science, Orhanl{\i} $-$ Tuzla, 34956, \.{I}stanbul, Turkey}
\author{K.\ Y.\ Ek\c{s}\.{i}}
\affiliation{\.{I}stanbul Technical University, Faculty  of Science  and  Letters,  Physics Engineering  Department,
  34469,  \.{I}stanbul, Turkey}
\author{E.\ G\"o\u{g}\"u\c{s}}
\affiliation{Sabanc{\i} University,
  Faculty of Engineering and Natural Science, Orhanl{\i} $-$ Tuzla, 34956, \.{I}stanbul, Turkey}
\author{T. G\"uver}
\affiliation{\.{I}stanbul University, Science Faculty, Department of Astronomy and Space Sciences, Beyaz{\i}t,
34119, \.{I}stanbul, Turkey}

\begin{abstract}
Aql~X--1 is one of the most prolific \acp{LMXBT} showing outbursts almost annually. We present the results of our spectral analyses of RXTE/PCA observations of the 2000 and the 2011 outbursts. 
We investigate the spectral changes related to the changing disk-magnetosphere interaction modes of Aql~X--1. 
The X-ray light curves of the outbursts of \acp{LMXBT} typically show phases of fast rise and exponential decay. 
The decay phase shows a ``knee'' where the flux goes from the slow decay to the rapid decay stage. 
We assume that the rapid decay corresponds to a weak propeller stage at which a fraction of the inflowing matter in the disk accretes onto the star. 
We introduce a novel method for inferring, from the light curve, the fraction of the inflowing matter in the disk that accretes onto the NS depending on the fastness parameter. 
We determine the fastness parameter range within which the transition from the accretion to the partial propeller stage is realized. 
This fastness parameter range is a measure of the scale-height of the disk in units of the inner disk radius. 
We applied the method to a sample of outbursts of Aql~X--1 with different maximum flux and duration times. 
We show that different outbursts with different maximum luminosity and duration follow a similar path in the parameter space of accreted/inflowing mass flux fraction versus fastness parameter.

\end{abstract}


\section{Introduction}
\label{intro}

\acp{LMXB} are systems containing an accreting compact object---a neutron star (\ac{NS}) or a black hole---and a low mass 
companion ($M_{\mathrm c} \lesssim 1\,M_{\odot}$). In these systems, the mass transfer mechanism is the Roche lobe overflow \citep{fra+02}. 
The transferred matter has angular momentum and forms an accretion disk rather than  directly infalling onto the compact 
object \citep{pri72}. Matter in the bulk of the disk rotates in Keplerian orbits and slowly diffuses inwards while the angular momentum is transported outwards by turbulent viscous processes 
\citep{sha73}. 

If the accreting object is a \ac{NS} the inner parts of the disk may be disrupted at a location beyond the stellar surface. The location of the inner radius of 
the disk, $R_{\rm in}$, is determined by the balance between the material and magnetic stresses in the disk \citep{gho79a,gho79b} which in turn  depend on the mass inflow rate $\dot{M}$ in 
the disk, the magnetic dipole moment $\mu_{\ast}$ and the spin angular velocity $\Omega_{\ast}$ of the star \citep{lam+73}.
Such magnetized star-disk systems \citep[see][for a review]{rom15} may show three different stages depending on the relation between the inner radius of the disk and the two other 
characteristic radii; the corotation radius $R_{\rm c}=(GM_{\ast}/\Omega_{\ast}^2)^{1/3}$ and the radius of the light cylinder $R_{\rm L} = c/\Omega_{\ast}$ \citep{lip92} where $G$ is the 
gravitational constant, $c$ is the speed of light and $M_{\ast}$ is the mass of the star: 
\begin{itemize}
\item The \textit{accretion} stage in which  $R_{\rm in}<R_{\rm c}$ resulting with most (if not all) of the mass flux in the disk to reach to the surface of the 
\ac{NS}. 
\item The \textit{propeller} stage \citep{ill75} in which $R_{\rm c} < R_{\rm in} \lesssim R_{\rm L}$ resulting with none (if not a small fraction) of the 
inflowing mass to reach the surface of the \ac{NS} due to centrifugal barrier formed by the rapidly rotating 
magnetosphere. 
\item The \textit{radio pulsar} stage in which $R_{\rm in}$ is even further away from the star, possibly larger than 
$R_{\rm L}$.
\end{itemize}
The gravitational potential energy of infalling material powers the \hbox{X--ray} luminosity:
\begin{equation}
L_{\mathrm X} = G M_{\ast} \dot{M}_{\ast} / R_{\ast}
\end{equation}
\citep{dav73} where $R_{\ast}$ is the radius of the \ac{NS}. Here $\dot{M}_{\ast}$ is the mass accretion rate onto the \ac{NS} and this may be different than the mass flow rate in the disk, $\dot{M}$,
in unsteady regime such as occurring during an outburst. 

\acp{AMXP} \citep{wij98} constitute a subclass of \ac{LMXB} systems which  show coherent pulsations in their X-ray light curves resulting from accretion 
onto the magnetic pole of a \ac{NS} from a  disk truncated by magnetic stresses \citep[see][for a review]{patruno}.   All \acp{AMXP} are transient systems showing
outbursts in their X-ray light curves.
A typical X-ray light curve of an outburst displays a fast rise and an exponential 
decay. Following the released energy during an outburst, spectral state transition is realized from the low--hard state
(lower luminosity, harder spectrum) to the high--soft state (higher luminosity, softer spectrum) and viseversa.
In the hard state the spectrum 	outweighed by hard/Comptonized with a soft/thermal component and
in the soft state the spectrum is more dominated by soft/thermal component \citet{lin+07}. 
Accordingly, $\dot{M}$ rises steeply and declines slowly during the outburst within several weeks while magnetic dipole moment and angular velocity of the 
neutron star are relatively constant. As $\dot{M}$ changes these systems may manifest the above mentioned stages of disk-magnetosphere interaction.
\acp{AMXP} may thus serve as a lab for exploring the transitions between these different stages.

The decay stage of the X-ray light curves of \acp{AMXP} show a `knee' marking the transition from a slow decline to a rapid decline stage \citep[e.g.][]{ibr09}.
The cause of this change in their light curve was suggested to be a transition from the accretion stage to the propeller stage \citep{zha+98pro,gil+98,cam+98,asai+13} as assumed in this work. Similar `knees' in the light curves  are seen in black hole binaries which do not have magnetic fields and can not show propeller effect. Such transitions are assumed in some other references to be due to thermal disk instability model \citep[see][for a review]{las01}. The pulsations of SAX J1808.4-3658 are detected even at very low luminosities at which the system would be expected to be well in the propeller stage \citep{men+99, ust+06, rom04}.
This may indicate that the ``propeller effect'' is not ideal but a fraction 
\begin{equation}
f \equiv \dot{M}_{\ast}/\dot{M}
\label{f}
\end{equation}
of the inflowing mass reaches the surface of the star \citep{asai+13, cam+01, cui+97}.
This fraction would be a function of the fastness parameter of the system
\begin{equation}
\omega_{\ast} \equiv \Omega_{\ast} / \Omega_{\mathrm K}(R_{\rm in}) = \left( R_{\rm in}/R_{\mathrm c} \right)^{3/2}
\label{fastness}
\end{equation}
\citep{els77} where $\Omega_{\mathrm K}=\sqrt{GM/R^3}$ is the Keplerian angular velocity in the disk. In the simplest picture of an ideal propeller surrounded by an infinitely thin disk, $f$ is a step function
\begin{equation}
f(\omega_{\ast}) = 
\begin{cases}
1, & \text{for }  \omega_{\ast} \leq 1 \\
0, & \text{for }  \omega_{\ast} > 1
\end{cases}
\qquad \mbox{ideal propeller.}
\label{f_ideal}
\end{equation}
In real disks with a finite scale-height, $H$, a regime of partial accretion may be 
realized in which $f$ is expected to change smoothly with $\omega_{\ast}$. 
This is because accretion can proceed from higher latitudes of the disk even while 
the disk midplane is propelled \citep{rom04,eks11}. 
Indeed, the disk may become thicker to allow for such accretion as the fastness parameter goes above unity, 
or transition to the propeller stage may be induced as a result of the evaporation and thickening 
of the inner disk \citep{gungor+14}. 
The smoothness of the transition will be 
a measure of $H(R_{\rm in})/R_{\rm in}$. Being a dimensionless function of a dimensionless 
parameter $f=f(\omega_{\ast})$ should be unique for different outbursts of a system. 
In general $f$ may also depend on the inclination angle between the rotation and magnetic axis and so may vary for 
different systems. Theoretical estimates for $f(\omega_{\ast})$ were presented by  \citet{lip76} for 
spherical accretion and by \citet{men+99} for 
the quasi-spherical disk accretion case. The latter authors showed 
that $f = (3/8)\omega_{\ast}^{-4}$ at the $\omega_{\ast} \gg 1$ limit. 
The general case was investigated by \citet{eks11} with an application to SAX~J1808.4--3658, another \acp{AMXP}. 
In the present work we attempt to extract $f=f(\omega_{\ast})$ from observations, for the first time in literature as to our 
knowledge.

Aql~X--1 is one of the most active \acp{LMXBT}, exhibiting about 25 outbursts from 1996 until 2016
\citep{gungor+17a, mesh+17, mai08, cam+13, asai+13}. Over 20 years of observations of Aql~X--1, 
the pulsations were observed only in a very short duration of 150~s
which makes the source is classified as an intermittent \ac{AMXP}.
The observed spin frequency of $\nu_{\ast}= 550.27$~Hz \citep{cas+08} and the observations of
Type-I X-ray bursts \citep{koy+81} firmly establish that the compact object in this system is a neutron star.

In this work we investigate \ac{RXTE/PCA} observations of 
Aql~X--1, 
to study the transition into and from the accretion regime to the propeller regime with the assumption of that
the outbursts of Aql X--1 happen as a result of viscous thermal instability.
In \autoref{obs}, we explain the details of data reduction procedures, 
and present their outcomes.
In \autoref{method}, we present the method we use to infer $f(\omega_{\ast})$ from the X-ray light curves.
In \autoref{app}, we apply the method to the outcomes of our observational analysis and
present the results.
Finally, in \autoref{discuss}, we discuss our results and conclude.

\section{Observation and Data Analysis}
\label{obs}

\begin{figure*}
\centering
  \includegraphics[angle=90, scale=0.35]{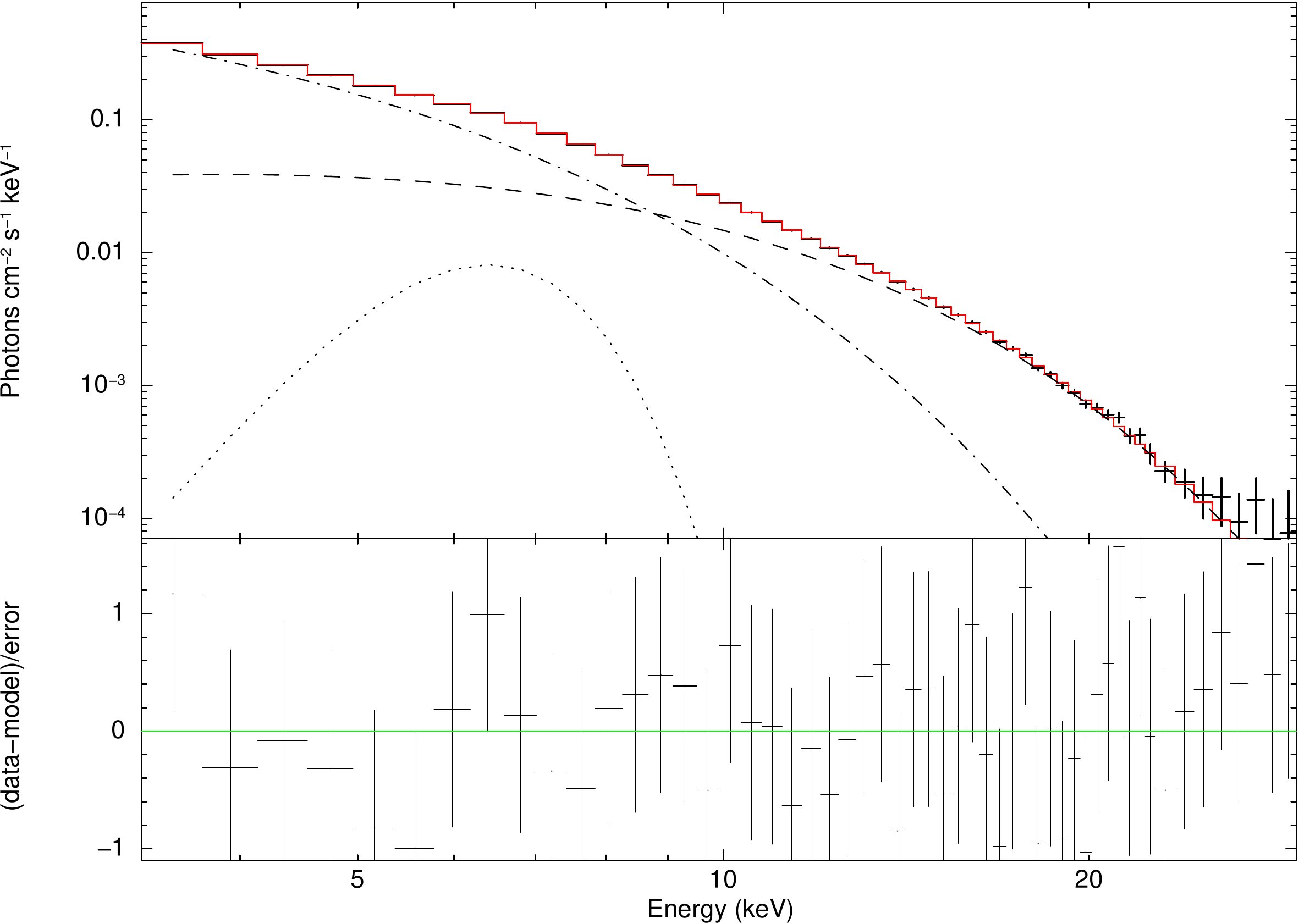}
  \includegraphics[angle=90, scale=0.35]{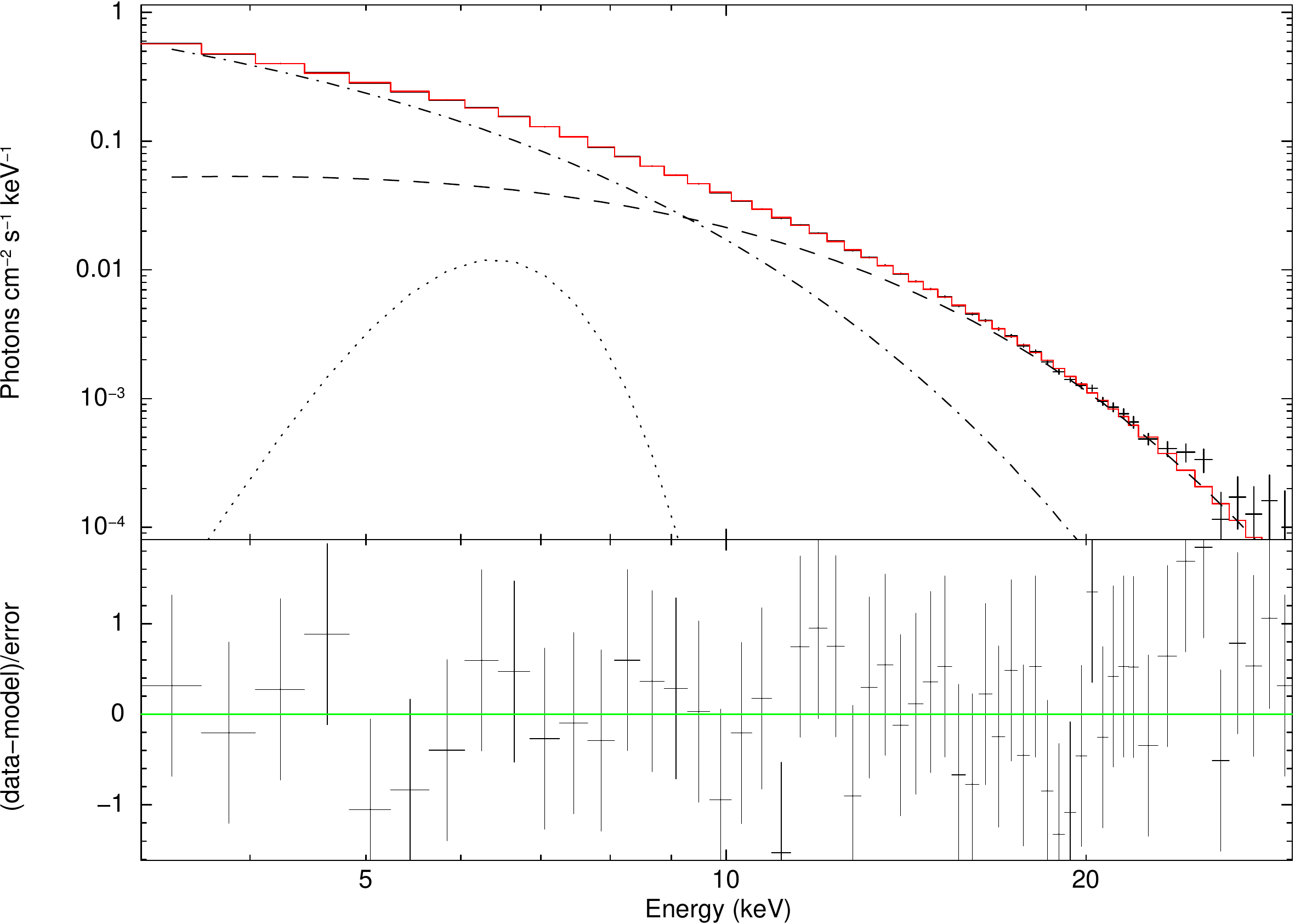}
     \caption{The X-ray spectrum of Aql~X--1 during outburst
     obtained by \ac{RXTE/PCA} in the  $3.0-30.0$~keV energy range
     during the 2000 (ObsID 50049-02-08-03, \textit{ the left panel}) and the 2011 (Obs ID 96440-01-05-01, \textit{the right panel}) outbursts, respectively.
     The best fits obtained by using the \textit{bb+diskbb+(ga)} model are shown with red line. 
     Lower panels show the residuals in terms of sigma.
     The dashed, the dashed dotted and the dotted curves show the blackbody,
     disk blackbody and the Gaussian components, respectively.}
         \label{fit-I}
\end{figure*}

The detected X-ray luminosity is assumed to originate from the \ac{NS} surface and the inner parts of the disk.
In order to separate these two components we performed a spectral analysis of \ac{RXTE/PCA} observations for the 2000 and  2011 outbursts
(52 and 51 pointed observations, respectively). We eliminated observations with thermonuclear bursts and those with very low S/N ratio.
The average exposure times of the selected observations $\sim$ 1800 s and $\sim$ 1450 s, respectively. 
These observations cover the entire durations of both outbursts, i.e.,\ the fast rising, the slow decay and the fast decay phases.

We analysed the \ac{RXTE/PCA} data using HEASOFT\footnote{https://heasarc.gsfc.nasa.gov/docs/software/lheasoft/} 
version 6.17. Since Aql~X--1 is a very
bright source in the X-ray band, we used only \acs{PCU}2 (\acl{PCU}) \acused{PCU} which was
always operational in all pointings. We generated the response files for each observation using PCARSP
version 11.7.1 and we used the latest module file\footnote{pca\_bkgd\_cmbrightvle\_eMv20051128.mdl}
for the background model. 

All spectra were modelled in the $3.0-30.0$~keV range using XSPEC package\footnote{An X-Ray Spectral Fitting Package v12.8.2,
https://heasarc.gsfc.nasa.gov/xanadu/xspec/}. 
We added a $1.0\%$ systematic error to the data during chi-squared test 
to take into account systematic instrumental uncertainties. We hypothesize two different models to represent the spectrum.
We  first modelled all spectra using a combination of blackbody, disk blackbody
and a Gaussian component to account for fluorescent iron emission. Second, we take the Comptonization into account with the assumption of
that the \ac{NS} surface and the Comptonization cloud have the same temperature.

\subsection{Model I: Blackbody assumption}
\label{model-I}

The blackbody component (\textit{bbody} in XSPEC) represents the X-ray emission
originating from the hot spot at the pole of the \ac{NS} fed by accretion. The disk blackbody
component (\textit{diskbb} in XSPEC) represents the X-ray contribution of the inner layers
of the accretion disk. After determining the best fit parameters for \textit{blackbody + disk blackbody}
model for the \ac{RXTE/PCA} data, a Gaussian line is added to represent the
iron line. 
We used a constant neutral hydrogen column density of
$N_{\rm H} = 3.4 \times 10^{21}$~atoms~cm$^{-2}$ \citep{mac03spec} using the model by \citet[\textit{phabs} in XSPEC]{bal92}. In \autoref{fit-I},
we show an example of the fits we performed for each data set. A typical normalisation value of diskbb component is around $(R_{\rm in}/D_{10})^2 \cos \theta = 100$ (for ObsID 50049-02-08-03) where $R_{\rm in}$ is the inner radius of the disk in km, $D_{10}$ is the distance in units of 10~kpc and $\theta$ is the viewing angle of the  disk, giving $R_{\rm in}= 4.5~{\rm km}/\sqrt{\cos\theta}$ assuming the source is at 4.5 kpc \citep{gal+08}.  This value of $R_{\rm in}$ is smaller than typical radius of a NS for a face-on disk ($\theta = 0$) and reaches reasonable values for $\theta \gtrsim 75^\circ$ corresponding to almost edge on view. The reason for small inner radius may also be an indication that Comptonization is significant \citep{lin+07} and the components are not fully separated as we assume. 

\begin{figure*}
\centering
  \includegraphics[angle=0, scale=1]{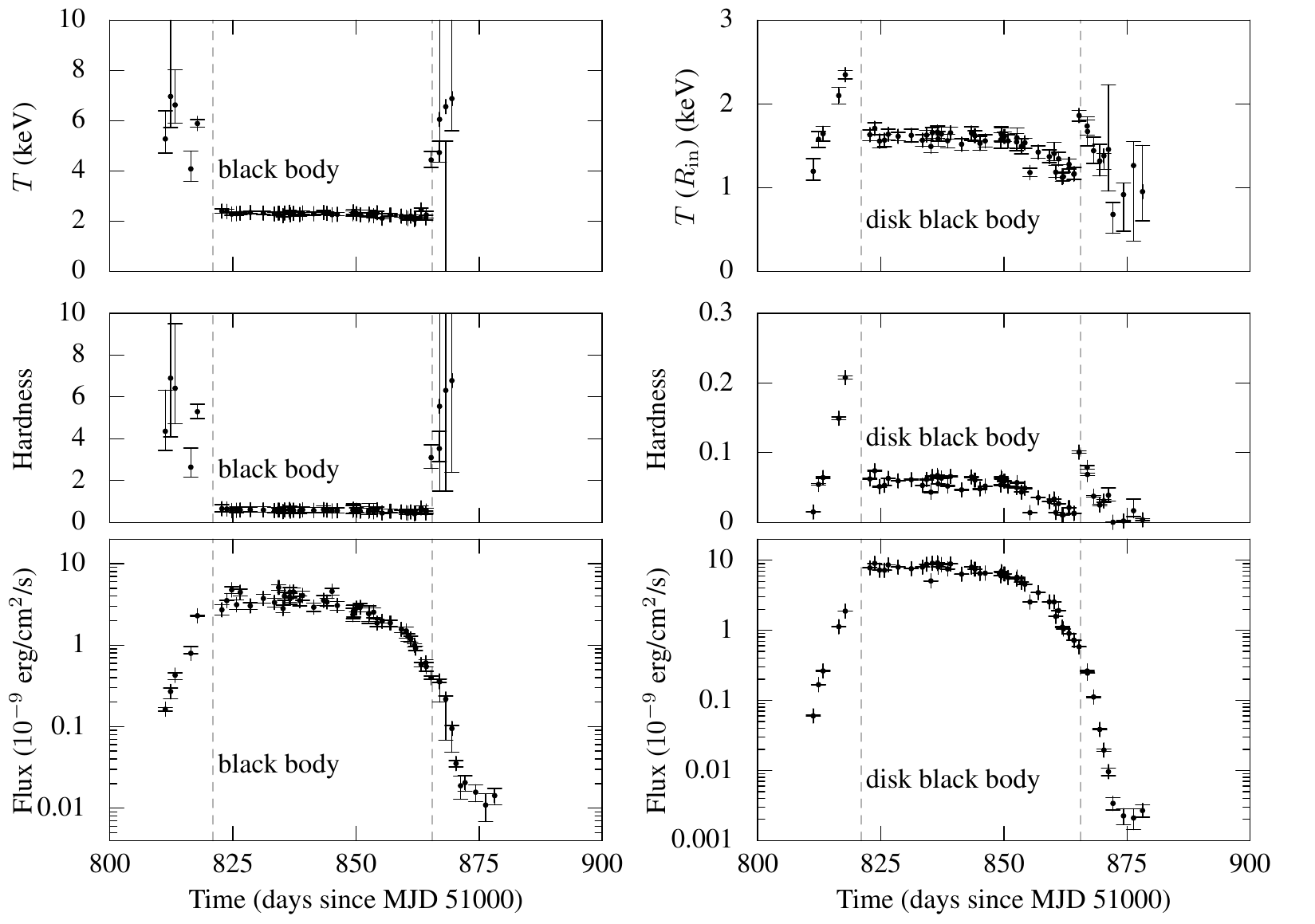}
     \caption{Evolution of spectral parameters during the 2000 outburst of Aql~X--1. 
     The top panels show the time evolution of the blackbody temperature (left) and the inner disk temperature of the blackbody (right). The middle panels 
     show the time evolution of the hardness parameter 
     only for the blackbody component (left) and
     only for disk blackbody component (right).
     The bottom panels show the evolution of the flux of the blackbody 
     component (left), and the disk blackbody component (right).
     The vertical lines show the times of the state transitions.}
         \label{bbdiskbb-2000}
\end{figure*}

\begin{figure*}
\centering
  \includegraphics[angle=0, scale=1.0]{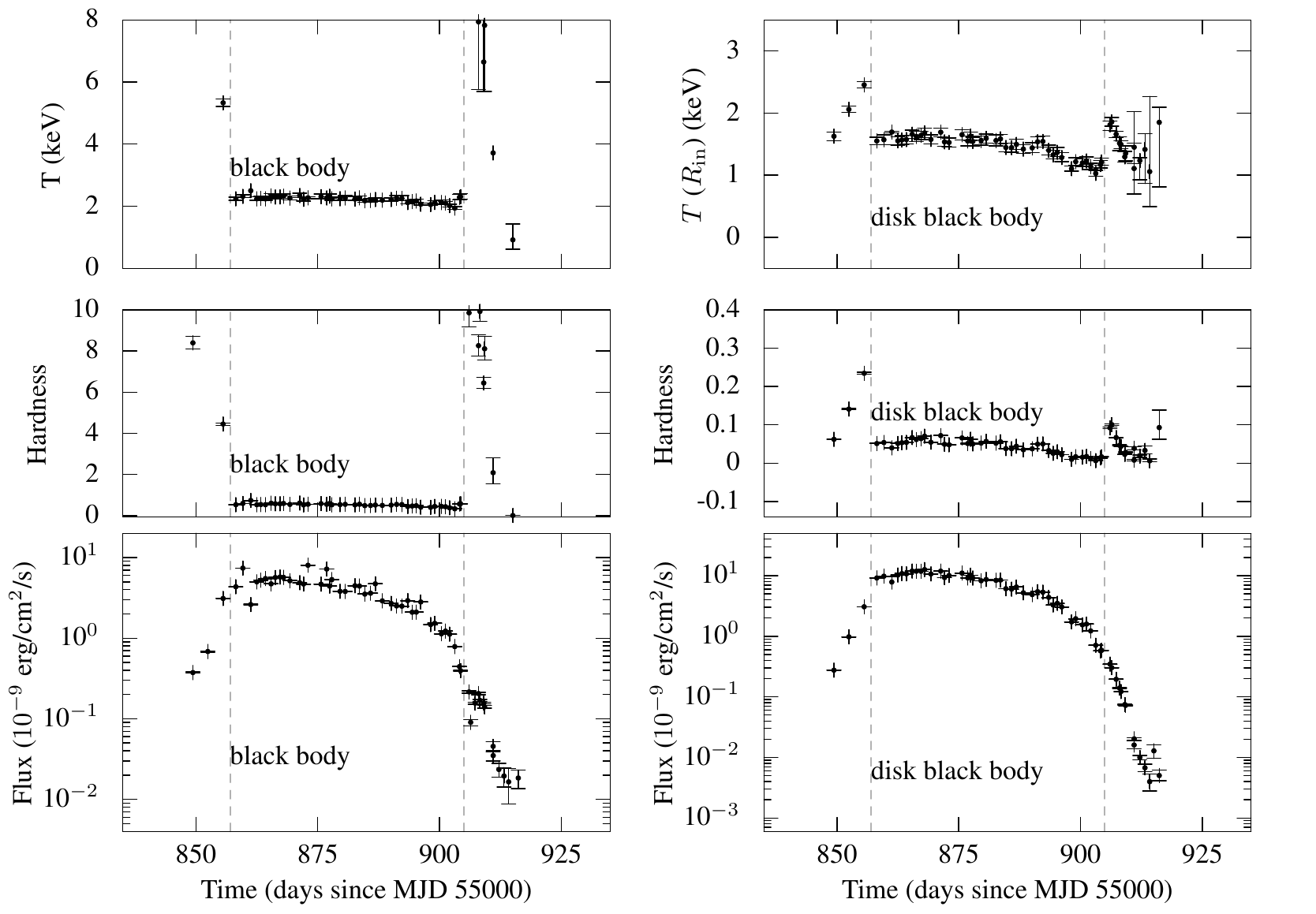}
     \caption{Same as \autoref{bbdiskbb-2000} but for the \ac{RXTE/PCA} data during the 2011 outburst of Aql~X--1.}
         \label{bbdiskbb-2011}
\end{figure*}

We calculated the unabsorbed fluxes for the best fit of the X-ray spectra using
\textit{blackbody + disk blackbody + Gaussian} model for the blackbody and the disk
blackbody components, separately  in the range of $3.0-30.0$~keV ($F_{3-30}$) for \ac{RXTE/PCA} data.
Bottom left and bottom right panels of \autoref{bbdiskbb-2000} and \autoref{bbdiskbb-2011}
show the light curves for each component 
for the 2000, the 2011 outburst, respectively.

Moreover, by taking the \textit{pivot} energy 
($E_{\rm pivot}$)---the energy in which we used to calculate hardness parameter---as $10$~keV,
we computed the fluxes for $3.0-10.0$~keV ($F_{3-10}$) and $10.0-30.0$~keV ($F_{10-30}$) energy ranges, 
following \citet{zha+98pro}. We then obtained the hardness evolution using the
ratio of $F_{10-30}/F_{3-10}$ (\autoref{bbdiskbb-2000} and \autoref{bbdiskbb-2011} middle left, middle right). The free parameters of the
blackbody and the disk blackbody are the temperature of the blackbody component ($T_{\rm bb}$), the inner
disk temperature of the disk blackbody component ($T_{\rm diskbb}(R_{\rm in})$) and the normalisations of
the models. We show the evolution of $T_{\rm bb}$ (top left) and $T_{\rm diskbb}(R_{\rm in})$ (top right) 
in keV in upper panels of
\autoref{bbdiskbb-2000} and \autoref{bbdiskbb-2011}.
We provide the final model parameters in \autoref{bbdiskbb.2000} and \autoref{bbdiskbb.2011}
for the 2000 and 2011, respectively.

In both components, we see that the spectral parameters evolve in a similar
way.
With the commence of the outburst $T_{\rm bb}$ drops down to $\sim 2$~keV while hardness ratio of the blackbody decreases to $\sim 1$ 
and both parameters stay stable during the bright phase of the outburst.
It must be mentioned that in the low--hard state, the temperatures throughout the model I are very high in which this shows the lack of the model I in low--hard state. 
The hardness decreases to the pre-outburst level before the outburst ends and the system passes to the quiescent stage.
This transition to the rapid-decay stage, rather than to the quiescent stage, 
marks the transition to the hard state from the
soft state \citep[see][for a BH study which is applicable to NSs]{rem+06}.

\citet{zhang98} argued that the
transition in the hardness ratio is caused by the \textit{propeller effect} \citep{ill75}.
These critical moments are used to identify the beginning and the terminal of slow decay stage which are also used as 
the fit range in our method described in \autoref{method}.

\subsection{Model II; Comptonised blackbody assumption}
\label{model-II}

Even though the blackbody assumption works in many cases, the upscattering of the photons by the coronal electrons must also be taken into account. 
We checked the role of Comptonization by modelling all spectra in the low--hard state and few examples in the high--soft state corresponding to accretion
using a similar combination as in model~I by adding a Comptonization model \citep[\textit{comptTT} in XSPEC,][]{titar95}
in which the Wien temperature of the Comptonization model
is linked to the temperature of the blackbody model under the assumption that photons from the \ac{NS} surface act as the seed for inverse Compton process.
As it is mentioned in \citet{comptt}, plasma temperature is  connected 
to the optical depth and fairly constrained. Thus, plasma temperature is set to a reasonable value, $15.0$~keV \citep{lin+07}.

The geometry is chosen as disk and the $\beta$ parameter of Comptonization is obtained from 
the optical depth ($\tau$) using  analytic approximation \citep[see][for details]{titar95}.
$\tau$ and the normalisation of the Comptonization model are free fit parameters.
For the X-ray contribution of the heated inner disk layers, we added the disk blackbody component just as in
model I. We added the Gaussian emission line component with the peak energy of 6.4 keV and maximum sigma of 1.0 keV
to account for Fe line emission.
The input parameters of the disk blackbody component to the fit 
are taken from the resulting fit of model~I to better constrain the Comptonization effect on the blackbody component and to check whether adding Comptonization
corrects the high blackbody temperatures in the low--hard state.

We provide resulting parameters of model~II analysis in \autoref{comptt.2000} and \autoref{comptt.2011}
for the 2000 and the 2011 outbursts, respectively.
Model~II is preferable to model~I for representing the $T_{\rm bb}$ better in the fast rising phase of the X-ray 
light curve which corresponds to low--hard state since model~I gives $T_{\rm bb}$ values unphysically high.
We tracked the time evolution of $\tau$  during outburst.
$\tau$ drops down to the zero level, following the trend of hardness, indicating that the system changes from 
low--hard state to high--soft state.
The variance of the parameters $\tau$ and hardness of the blackbody component implies that Comptonization becomes 
ineffective in the high--soft state.
It is possible to interpret this simply by suggesting that the contribution of Comptonization during the accretion 
becomes harder to determine when the disk is closer the star and has enhanced contribution to the total X-ray flux.
The result implies that the blackbody 
component is sufficient for representing the soft--high. 
We will then assume that the light curve of the blackbody component 
represents the time evolution
of the luminosity of accretion onto the \ac{NS} surface.

\section{The partial accretion  regime of outbursts}
\label{method}

Here, we propose a simple method for extracting $f(\omega_{\ast})$ from the light curve, $L_{\mathrm X}(t)$. 
The method is based on the following assumptions:
\begin{itemize}
\item The rapid decay stage is a consequence of transition of the system to the propeller stage $\omega_{\ast} > 1$ 
and is not due to irradiation or any other process.
\item The transition to the propeller stage occurs because material stress declines with the accretion rate
hence is balanced by the magnetic stresses at a larger distance, now further away from the corotation radius. 
We note here that BH systems also may have receding inner disk radii possibly as a result of disk evaporation \citep[see e.g.][]{liu+99,mey+00}.
If the same mechanism also works in NS systems a different analysis than ours has to be employed.
\item The decay of the mass inflow rate $\dot{M}$ continues its evolution with no modification upon 
the transition of the system from accretion to  the propeller stage though a smaller fraction of it 
can now accrete onto the star leading to the appearance of rapid decline. 
This is possibly because of the delay in transferring the information of the changed inner boundary condition
 to the outer parts of the disk which keep transporting matter in.
\item The angular velocity $\Omega_{\ast}$ and magnetic moment $\mu_{\ast}$ of the neutron star does not change significantly during an outburst.
\end{itemize}

We also assume
$f=1$ (meaning that $\dot{M}_{\ast}=\dot{M}$) in the slow decay stage before the knee though there is reason to believe that some of the matter donored by the companion is ejected from the disk by winds on the way to the innermost disk. 
Soon after the maximum is reached the disk establishes a quasi-equilibrium stage which evolves self-similarly \citep{lyu87,lip02,sul08} where the mass flux will evolve as:
\begin{equation}
\label{luminosity}
\dot{M}(t) = \dot{M}_{0} \left(1 + \frac{t-t_0}{t_{\nu}} \right)^{-\alpha}.
\end{equation}
Here $t_{\nu}$ is the time-scale of the outburst decay (viscous timescale) and $ \dot{M}_0$ is the mass flux at $t_0$ which is the moment power-law decline starts. 
In the full accretion regime ($f=1$) the luminosity follows this trend so that we can fix $\dot{M}_0 = L_0 R_{\ast}/GM_{\ast}$
where $L_0$ is the luminosity at the moment of $t_0$. 
The value of the power-law index $\alpha$ depends
on the  pressure and opacity prevailing in the disk \citep{can+90,eks11}.
We have fixed $\alpha=1.25$ appropriate for a gas pressure dominated disk with bound-free opacity \citep{can+90}. Although the inner parts of the disk for high accretion rates will be dominated by radiation pressure and electron scattering opacity, the mass flux throughout the disk is regulated by the outer parts where gas-pressure and bound-free opacity dominates.


Aql~X--1 is a $\nu_{\ast}=\Omega_{\ast}/2{\rm \pi} = 550.27$~Hz AMXP  \citep{cas+08}.
The corotation and the light cylinder radii for a neutron star with this the spin frequency are  
$R_{\rm c}=(GM_{\ast}/\Omega_{\ast}^2)^{1/3}=2.5 \times 10^6$~cm (for $M_{\ast}=1.4~{\rm M}_{\odot}$) and $R_{\rm L}=c/\Omega_{\ast}=8.7 \times 10^6$~cm, respectively. 
The maximum critical fastness parameter above which the inner radius goes beyond the light cylinder for this system is $\omega_{\ast \max}=(R_{\rm L}/R_{\rm c})^{3/2} \simeq 6.5$.

The inner radius of the disk is proportional to the Alfv\'en radius, $R_{\rm in} = \xi R_{\rm A}$, where $\xi$ is a constant of order unity and generally taken as $\xi=0.5$ \citep{gho79a,gho79b}).
The Alfv\'en radius \citep{dav73} is
\begin{equation}
R_{\rm A} = \left( \frac{\mu^2}{\sqrt{2GM} \dot{M}} \right)^{2/7}
\end{equation}
where $\mu$ is the magnetic dipole moment of the star.  This designation could be valid only if it does not yield an inner radius smaller than the radius of the \ac{NS}. The disk could extend to the surface of the star only at the peak of brightest outbursts. As the Alf\'en radius scales with the mass flux as $R_{\mathrm A} \propto \dot{M}^{-2/7}$ which then implies, by \autoref{fastness}, that $\omega_{\ast} \propto \dot{M}^{-3/7}$ or rather
\begin{equation}
\omega_{\ast} = (\dot{M}/\dot{M}_{\rm c})^{-3/7}
\label{fast2}
\end{equation}
where $\dot{M}_{\rm c}$ is the mass inflow rate that would place the inner radius on the corotation radius and is related to $L_{\rm c}$, critical luminosity at which partial accretion starts, as $\dot{M}_{\rm c} = L_{\rm c} R_{\ast}/GM_{\ast} f_{\rm c}$ where $f_{\rm c}$ is the fraction of mass flux at this critical stage. 
This corresponds to the luminosity at which rapid decline commences.
The inner radius of disk in the quiescent stage at which $\dot{M}=0$ was formalized by \citet{ozs+14} in the context of a putative supernova debris disk 
around the Vela pulsar assumed to be in a strong propeller stage. 
In the decay stage that we consider here for Aql~X--1 we have already assumed mass keeps inflowing even for $\omega_{\ast}>1$ and the system is in a weak propeller regime so we find the scaling of the Alfv\'en radius appropriate.

In the propeller stage we assume that the mass inflow rate in the disk determining 
the inner radius continues with the same trend. Thus if all this inflowing matter 
could accrete we would have a luminosity continuing with the same trend of the accretion stage with no knee. 
The presence of the knee is assumed to be a consequence of partial accretion in the propeller stage: thus a fraction $f$ of $\dot{M}$ can accrete.
The rest of the material may be ejected from the system completely via jets \citep{tud+09}.
In order to describe this partial accretion we replace the fraction $f(\omega_\ast)$ given in  \autoref{f_ideal} for the ideal propeller with a smoothed step function that varies from unity to $f_{\min}<1$
\begin{equation}
f = \frac12 \left[ 1 + f_{\min} + (1-f_{\min}) \tanh \left(\frac{\omega_{\rm c} - \omega_\ast}{\delta}\right) \right]
\label{step}
\end{equation}
where $\omega_{\rm c}$ ($= 1$) is the critical fastness parameter at which the transition between accretion and propeller stages, and $\delta$ is  a measure of the
abruptness of this transition.

We first fit the region between the maximum of the light curve and the knee using \autoref{luminosity} 
to determine $t_{\nu}$, $L_0$, $t_0$ and $t_{\rm knee}$.
We then fit the light curves from the maximum of the outburst to the end of the data using $f(t)=f[\omega_{\ast}(t)]$ with the initial fit values. From the latter fit we obtained the values of $f_{\min}$ and $\delta$ which are the free fit parameters.

The above analysis assumes that the X-ray luminosity totally originates from accretion onto the \ac{NS}. For weakly magnetized neutron stars, such as Aql X--1, the inner radius of the disk is close to the star and the inner disk may contribute to the X-ray luminosity.
To obtain the luminosity due to accretion onto the star alone one needs to use spectral analysis (in \autoref{obs}).

We note that there are diverging views of how the propeller stage is realized. \citet{dan10} argued that propeller stage at low accretion rates will be realized by accumulation of matter  at the inner disk rather than being ejected out of the system. This will lead to bursts 
of enhanced accretion \citep{spr93,dan11} and steady state quiescent (dead) disk solutions \citep{sun77}. The timescales for the bursts is the viscous time-scale at the inner disk ($\tau_{\nu} \sim 1~{\rm ms}$) which is not resolved in the data we employ in this work. 
Whether the matter is ejected or is retained in the disk, the accretion rate onto the star and hence X-ray luminosity will decline and, therefore, from the point of view of the present work analysing the fraction of accreting matter onto the star depending on the fastness parameter remain to be relevant.

\section{Applications}
\label{app}

A broad classification of the outbursts of Aql X--1 is presented by \citet{gungor+14} based on the maximum flux and the duration of 
the outbursts. Accordingly, Aql X--1 shows three main types of outbursts: \textit{(i)} The \textit{long-high} outbursts
with outburst duration of $50-60$~days and a maximum flux of $37-61$~cnt/s.
\textit{(ii)} The \textit{medium-low} outbursts with $40-50$ days and a maximum flux of $13-25$~cnt/s.
\textit{(iii)} The \textit{short-low} outbursts with approximately 20 days duration 
time and reaching a maximum flux of $17-25$~cnt/s.

As we focus on the energetic outbursts, after performing the spectral analysis to the \ac{RXTE/PCA} data of the 2000 and 
the 2011 outbursts --both belonging to the long-high class-- and obtaining the light 
curves only for blackbody components explained in \autoref{model-I}, we applied the 
procedure described in \autoref{method} to calculate the fraction of mass flux reaching the \ac{NS} in the propeller stage.
In addition, to check the possible differences, we applied the technique to \ac{ASM} data of the 2000 outburst, 
\ac{MAXI} data of the 2013 outburst and the most energetic one, 2016 outbursts \citep{gungor+17a}.

\begin{figure*} 
\centering
  \includegraphics[angle=0, scale=1.0]{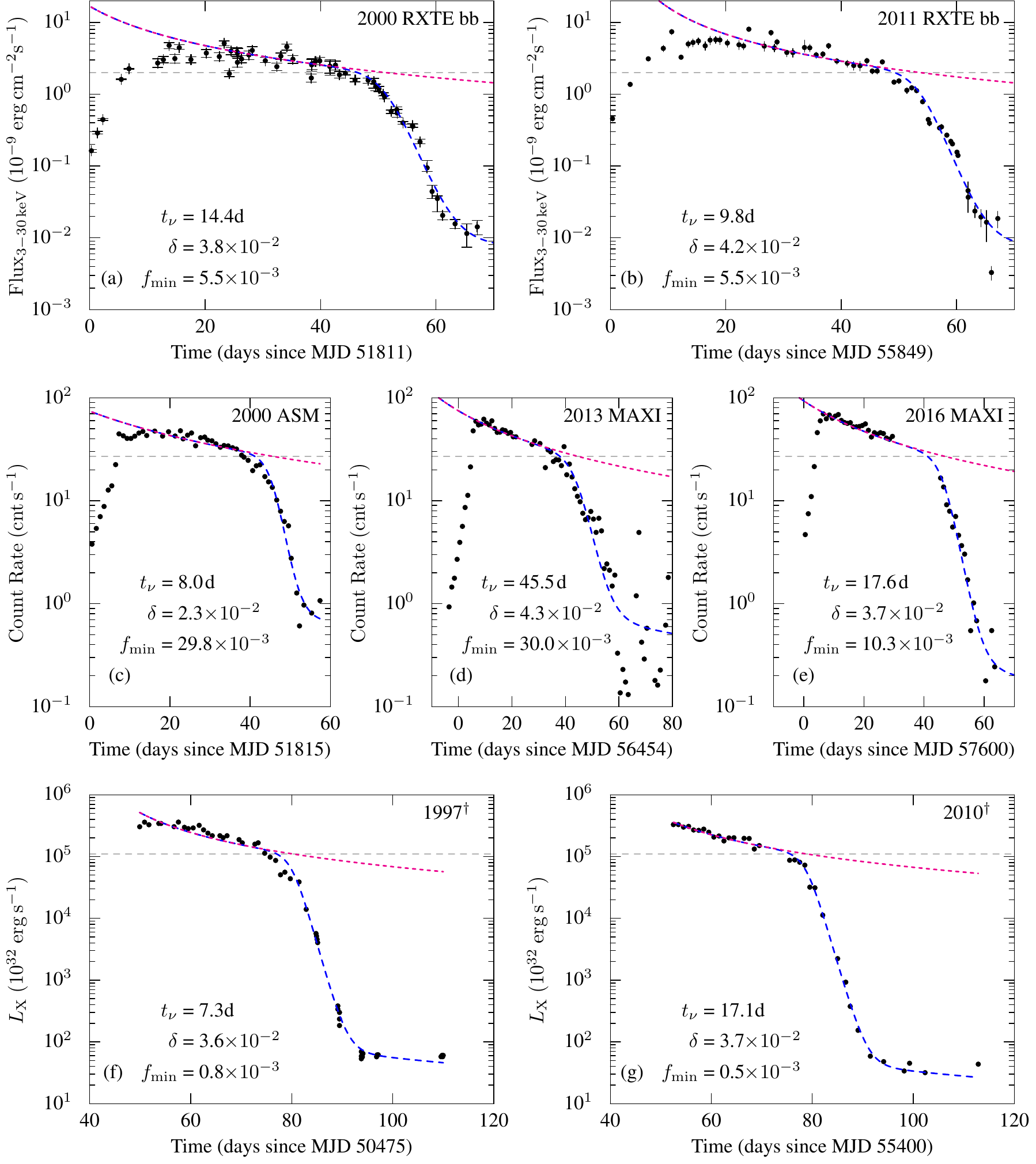}
  \caption{The X-ray light curves of the blackbody components of the \ac{RXTE/PCA} data of the 2000 (a) 
  and the 2011 (b) outbursts.
  The \ac{ASM} light curve of the 2000 outburst (c). The \ac{MAXI} light curves of the 2013 (d) and the 2016 (e) outbursts (Count rates are calibrated to \ac{ASM} level).
  The horizontal lines show the $L_c$ levels in \autoref{luminosity}.
  The pink curves show the best fits of \autoref{luminosity} between maximum of outbursts and the knee. The blue curves show the best fits to the total data.\\
  $^{\dagger}$The X-ray light curves of the 1997 (f) and the 2010 (g) outbursts obtained from \citet{cam+14}.}
         \label{fit-pure}
\end{figure*}

The decay stages of the 1997 and the 2010 outbursts of Aql~X--1 have been studied by \citet{cam+98} and \citet{cam+14}.
In the latter work the authors concluded that the rapid decay stage is likely caused by the transition to the propeller stage. 
We, also, applied the method to these outbursts to investigate the mass transfer characteristic for different classes.
In \autoref{fit-pure}, we show the light curves of all outbursts together with the fit function of \autoref{luminosity}.

\begin{figure} 
\centering
  \includegraphics[angle=0, scale=0.9]{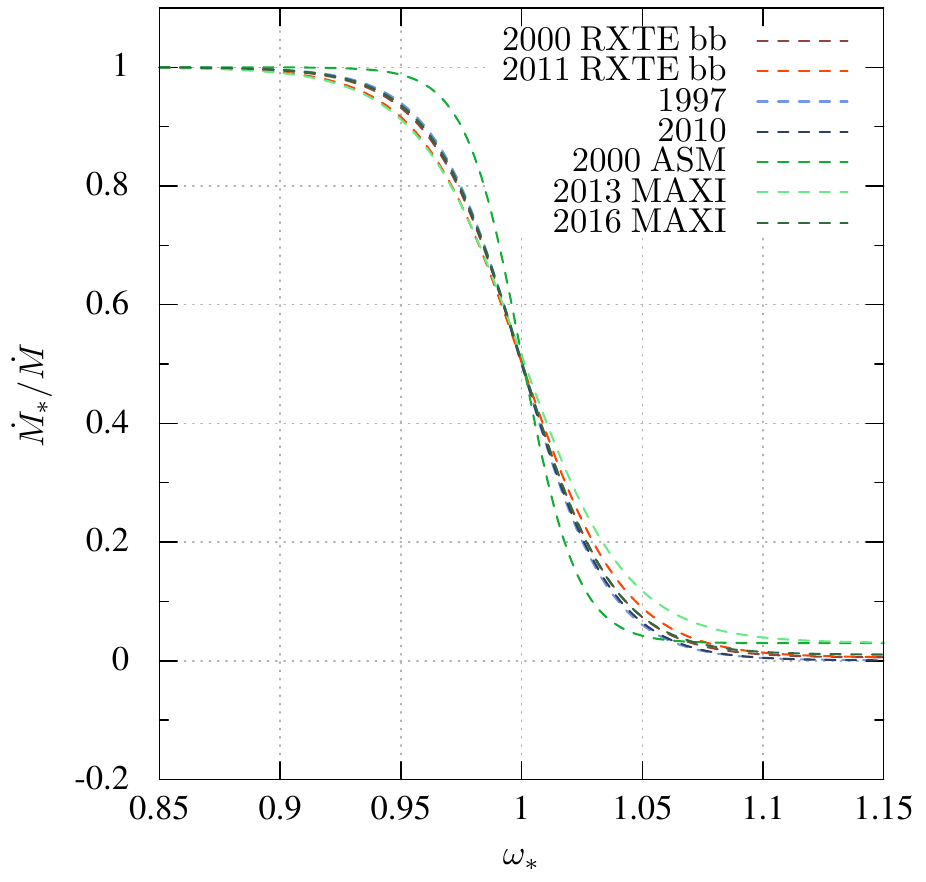}
     \caption{$f = \dot{M}_{\ast}/\dot{M}$ vs the fastness parameter $\omega_{\ast}$ relation obtained from the outbursts of Aql~X--1.
}
         \label{frac-all}
\end{figure}

In \autoref{frac-all}, we show  $f\equiv \dot{M}_{\ast}/\dot{M}$ vs $\omega_{\ast}$ i.e.\ \autoref{step} for all outbursts in our sample.
The numerical values of the parameters $f_{\min}$ and $\delta$ of \autoref{step} and $t_{\nu}$ of \autoref{luminosity} obtained by fitting the lightcurve of each outburst (as described in \S~\ref{method}) are given in \autoref{fit-model}.

\begin{table} [h]
\caption{Resulting values of the free parameters of the method.}
\begin{center}
\begin{tabular}{cccccccccccccccccccccc}
\hline
\hline
    & $t_{\nu}$ (days) & $f_{\min}~(10^{-3})$    &    $\delta~(10^{-2})$ \\
    \hline
2000 RXTE    &    14.4    &    5.5    &    3.8    \\
2011 RXTE    &    9.8    &    5.5$^{\dagger}$    &    4.2    \\
2000 ASM    &    8.0    &    29.8    &    2.3    \\
2013 MAXI    &    45.5    &    30.0$^{\dagger}$    &    4.3    \\
2016 MAXI    &    17.6    &    10.3    &    3.7    \\
1997$^{\ddagger}$    &    7.3    &    0.8    &    3.6    \\
2010$^{\ddagger}$    &    17.1    &    0.5    &    3.7    \\
\hline
\end{tabular}
\end{center}
\footnotesize{
\begin{flushleft} $^{\dagger}$The values are fixed to the outcome of last outburst since there is no enough data to obtain base level of the step function. \\
$^{\ddagger}$The light curve data is taken from \citet{cam+14}.
\end{flushleft} }
\label{fit-model}
\end{table}

\section{Discussion}
\label{discuss}

In this paper, we studied the evolution of the disk-magnetosphere interaction from accretion to the propeller regime
using the X-ray data of \hbox{Aql~X--1}.

We analysed the RXTE/PCA data of the 2000 and the 2011 outbursts of Aql~X--1.
We modelled each spectra using a combination of
blackbody and disk blackbody and obtained the lightcurves and time evolution of the spectral parameters for each component  separately. 

The blackbody component represents 
seed photons released at the hot spot on the NS poles while Comptonization plays a role in the spectral formation. 
We checked the validity of the blackbody assumption using a Comptonization model \citep[\textit{comptTT} in XSPEC,][]{titar95} 
with a linked Wien temperature parameter to the blackbody 
temperature and showed that
Comptonization is not significantly dominant during the slow decay stage. 
This supports the view of that X-ray flux due to
accretion is represented by the blackbody radiation. 
Different models could be employed to separate the contributions of different components 
in the total X-ray flux (\citealp[e.g. \textit{nthcomp(blackbody)+diskbb},][]{sak+12}; \citealp[][]{lin+07}).

Although Comptonization is ineffective in the soft state, 
adding it to the model corrects the blackbody temperatures at the hard state. 
Temperatures inferred without employing Comptonization are too high to be explained via known physical processes 
(Obs~\# 1--5 in \autoref{bbdiskbb.2000} and Obs~\# 1--3 in \autoref{bbdiskbb.2011}). The last two data of the hard state, however, 
show similarly high temperatures even with the Comptonization component included.
These data correspond to the episode before the maximum of the X-ray light curve
and the spectra at this episode is still hard while luminosity is high.
This may indicate to a \textit{high--hard interstage} which occurs at the rising stage of the outburst before the low--hard stage and high-soft stage at the decay stages.
Considering that a similar situation exists in transitions from high-soft stage to low--hard stage, we suggest that the system shows
a \textit{pre-propeller} stage at the similar luminosity level while the disk drifts to the inner layers.
As the rising phase of the X-ray light curve is shorter than the decay phase,
the pre-propeller stage lasts shorter than the propeller stage.

Recently, \citep{tsy+17} suggested that the decline in X-ray flux during the outbursts of high mass X-ray binary systems (containing slowly rotating neutron stars) is due to transition to a cold (neutral; $T \lesssim 6500~{\rm K}$) disk state. In this case, accretion rate is low due to the suppression of magnetorotational instabilities. In the case of Aql~X--1, the rapid rotation and relatively small magnetic field of the compact object will likely eliminate the possibility of accretion from a cold disk \citep[see Figure~3 of][]{tsy+17}.

For probing the transitions from the accretion to the propeller stage,
we introduced a novel method to
determine the ratio of the mass accretion rate onto NS pole to the mass inflow rate at the inner disk, depending on the fastness parameter, from the observed X-ray light curve.
The method depends on the assumptions that the rapid decay stage corresponds to the propeller stage at which partial accretion proceeds and that the X-ray flux is due to 
weak accretion onto the poles. Our results imply a more steeper decline of $f$ with the fastness parameter than predicted by the existing theoretical models for spherical accretion \citep{lip76} and quasi-spherical accretion \citep{men+99}.

Herein, the outcome of the analysis of the X-ray data of the 2000 and 2011
outbursts allows to better represent the emission from the hot spots on the NS as the result of accretion.
By using only the blackbody light curves for these outbursts,
we transform the flux vs time data
to the fraction ($f \equiv \dot{M}_{\ast}/\dot{M}$) 
vs fastness parameter ($\omega_{\ast}$) domain. Comparing all  outbursts in this domain 
find that different outbursts with different time scales and maximum fluxes follow a similar path in $f-\omega_{\ast}$.
The step like function representing the outbursts in $f$ vs $\omega_{\ast}$ space is a 
function constrained in a narrow band for this system (Aql X--1). The abruptness, $\delta$, of the step function given 
in \autoref{step} must be related to the thickness of inner layers of 
the disk (disk scale-height $H$) in units of inner disk radius, and the inclination angle between the rotation and magnetic dipole axis of the \ac{NS}. 

The transition to the propeller regime allows for an estimate of the magnetic field of the NS \citep{cam+98,dis03,asai+13,cam+14,muk+15,kin+16}. Our continuous representation of the fraction of the accreting mass flux indicates that $f_{\rm c}=0.5$ (see \autoref{frac-all})  when $R_{\rm in}=R_{\rm c}$ ($\omega_{\rm ast}=1$). This leads to an higher estimate of the magnetic dipole moment which is $\sqrt{2}\simeq 1.5$ times greater than the previous estimates which assume $f=1$ when at this stage. We find that
\begin{align}
\mu = 8.0\times & 10^{26}\, {\rm G~cm^3}\, M_{1.4}^{1/3} R_{10}^{1/2} \nonumber \\ 
      & \left( \frac{\xi}{0.5}\right)^{-7/4} \left( \frac{L_{\rm c}}{1.1\times 10^{37}~{\rm erg~s^{-1}}} \right)^{1/2}  
\end{align}
where $M_{1.4}$ is the mass of the NS in units of 1.4 solar mass and $R_{10}=R_{\ast}/10^6\, {\rm cm}$. Here we assumed $L_{\rm c}=1.1\times 10^{37}~{\rm erg~s^{-1}}$ which we inferred from the ``knee'' seen in the lightcurves of the 1997 and 2010 outbursts in \autoref{fit-pure}.

Our assumption that the rapid decay stage corresponds to the weak propeller regime is not commonly accepted. Given the existence of black hole systems which also  
show a similar knee in the lightcurve, it may be argued that the rapid decay stage is a property of the disk instability model underlying the outburst \citep{kin98,cam+14}. 
According to this picture
a cooling front moving inwards in the disc is the cause of the transition to the rapid decay stage. 
The possibility that the knee seen in the X-ray light curves of outbursting systems is due to the transition to the propeller stage as well as to the disk instability, 
is argued in the literature \citep[e.g.][]{gil+98, ibr09, eks11,asai+13,gungor+14}. 
The narrow range of $\delta$ values  characterizing the transition, as we inferred in this work, supports the view that the rapid decay stage represents transition to the partial accretion regime. The wide rage of the $f_{\min}$ values may then indicate the varying contribution of sources other than accretion (such as cooling of the star) to the quiescence luminosity.
This would imply that that the rapid decay stage in black hole systems has a different cause than neutron star systems such as the truncation of the disk.

\section{Conclusions}

Based on our model and the related investigation we conclude that:
\begin{itemize}

\item The range $\omega_{\ast} \lesssim 0.9$ is the slow decay stage. All of the material 
transferred from outer disk accretes onto the  NS. As $\dot{M}$ ($=\dot{M}_{\ast}$ in accretion stage) 
decreases in time, the luminosity declines (slow decay stage) while $R_{\rm in}$ expands back to $R_{\rm c}$.

\item The range $0.9 \lesssim \omega_{\ast}\lesssim 1.1$ is the partial accretion regime. A fraction of inflowing material to the inner layers of the disk may transfer onto the NS.
The rest may be thrown to outer layers of the disk or expelled from the system via jet mechanism. 

\item The range $\omega_{\ast} \gtrsim 1.1$ 
is the fully developed propeller stage and the neutron star may even act as an isolated \ac{NS} \citep{eks05}.

\end{itemize}


\section*{Acknowledgements}

We gratefully acknowledge the anonymous referee for very constructive comments and suggestions.
CG thanks The Scientific and Technological Council of 
TURKEY (TUBITAK) for the 2214-A Scholarship. 
CG is grateful to Prof. Neuh\"auser for 1-year research project in Jena/Germany.
CG appreciate Prof. Santangelo for T\"ubingen visit and useful discussion.
KYE, TG and CG acknowledge support from TUBITAK with the project
number 112T105. KYE and CG thank \.Istanbul Technical University Scientific Research
Projects Unit (ITU-BAP) for the support with the project number of 38339.
This research has made use of the XRT Data Analysis Software (XRTDAS) developed under the responsibility
of the ASI Science Data Center (ASDC), Italy.


\begin{table*}
\caption{Best fit parameters of \textit{blackbody + disk blackbody + Gauss} model for the 2000 outburst.
The horizontal lines in the table indicate the state transitions according to hardness parameter of the blackbody component.}
\begin{center}
\tiny
\begin{tabular}{cccccccccccccccccccccc}
\hline
\hline
Obs \#	&	ObsID	&	MJD$-51000$	&	$kT_{bbody}$					&$	 kT_{diskbb}					$&	$\chi^2/d.o.f.^{a}$	&	Hardness$^{b}$					&	Hardness$^{b}$					&	Flux$^{c}_{bbody}$					&	Flux$^{d}_{diskbb}$					\\
	&		&	(days)	&	(keV)					&	(keV)					&		&	bbody					&	diskbb					&						&						\\ \hline
1	&	50049-01-03-00	&	811.28564	&$	5.27	\pm{			1.12	}$ & $	1.19	\pm{			0.15	}$ & $	0.90	$&$	4.36	\pm{			1.97	}$ & $	1.50	\pm{			0.06	}$ & $	1.63	\pm{	0.08			}$ & $	0.60	\pm{			0.01	}$\\
2	&	50049-01-03-01	&	812.34953	&$	6.96	\pm{	1.23			}$ & $	1.58	\pm{	0.10			}$ & $	1.14	$&$	6.89	\pm{	2.81			}$ & $	5.45	\pm{			0.13	}$ & $	2.70	\pm{	0.52			}$ & $	1.68	\pm{			0.02	}$\\
3	&	50049-01-03-02	&	813.27735	&$	6.63	\pm{			1.40	}$ & $	1.64	\pm{			0.09	}$ & $	1.00	$&$	6.41	\pm{	1.69			}$ & $	6.43	\pm{			0.13	}$ & $	4.27	\pm{	0.49			}$ & $	2.63	\pm{			0.03	}$\\
4	&	50049-01-04-00	&	816.46667	&$	4.08	\pm{	0.71			}$ & $	2.10	\pm{			0.10	}$ & $	1.13	$&$	2.64	\pm{			0.92	}$ & $	14.93	\pm{			0.20	}$ & $	7.95	\pm{			1.67	}$ & $	11.29	\pm{			0.08	}$\\
5	&	50049-01-04-01	&	817.79791	&$	5.88	\pm{	0.16			}$ & $	2.35	\pm{			0.05	}$ & $	1.24	$&$	5.29	\pm{			0.36	}$ & $	20.77	\pm{			0.25	}$ & $	22.94	\pm{			0.40	}$ & $	18.80	\pm{			0.11	}$\\
\hline
6	&	50049-01-04-04	&	822.77448	&$	2.40	\pm{	0.08			}$ & $	1.63	\pm{	0.07			}$ & $	0.69	$&$	0.65	\pm{			0.19	}$ & $	6.23	\pm{			0.07	}$ & $	27.16	\pm{			4.01	}$ & $	77.91	\pm{	0.44			}$\\
7	&	50049-01-05-00	&	823.76888	&$	2.38	\pm{	0.07			}$ & $	1.71	\pm{	0.07			}$ & $	0.57	$&$	0.64	\pm{	0.06			}$ & $	7.40	\pm{	0.14			}$ & $	35.20	\pm{			0.20	}$ & $	90.30	\pm{	0.50			}$\\
8	&	50049-01-05-01	&	824.76115	&$	2.28	\pm{	0.05			}$ & $	1.56	\pm{	0.08			}$ & $	0.38	$&$	0.56	\pm{			0.11	}$ & $	5.15	\pm{	0.07			}$ & $	47.76	\pm{	4.49			}$ & $	72.13	\pm{	0.48			}$\\
9	&	50049-01-05-02	&	825.75496	&$	2.29	\pm{			0.08	}$ & $	1.57	\pm{	0.08			}$ & $	0.63	$&$	0.57	\pm{			0.16	}$ & $	5.31	\pm{	0.06			}$ & $	31.39	\pm{			4.32	}$ & $	71.95	\pm{	0.44			}$\\
10	&	50049-02-01-00	&	826.51291	&$	2.31	\pm{			0.05	}$ & $	1.64	\pm{	0.07			}$ & $	0.39	$&$	0.59	\pm{			0.12	}$ & $	6.31	\pm{	0.07			}$ & $	44.55	\pm{	4.34			}$ & $	86.14	\pm{	0.52			}$\\
11	&	50049-02-02-00	&	828.53443	&$	2.34	\pm{	0.05			}$ & $	1.61	\pm{	0.05			}$ & $	0.75	$&$	0.61	\pm{			0.12	}$ & $	5.96	\pm{	0.06			}$ & $	30.32	\pm{			3.03	}$ & $	79.80	\pm{	0.45			}$\\
12	&	50049-02-03-01	&	831.20242	&$	2.32	\pm{			0.07	}$ & $	1.62	\pm{	0.08			}$ & $	0.49	$&$	0.59	\pm{			0.15	}$ & $	6.12	\pm{	0.07			}$ & $	37.45	\pm{	4.64			}$ & $	75.74	\pm{	0.46			}$\\
13	&	50049-02-03-00	&	833.45856	&$	2.28	\pm{			0.07	}$ & $	1.57	\pm{	0.07			}$ & $	0.81	$&$	0.57	\pm{			0.15	}$ & $	5.30	\pm{	0.06			}$ & $	33.46	\pm{			4.39	}$ & $	79.95	\pm{	0.47			}$\\
14	&	50049-02-04-00	&	834.30025	&$	2.31	\pm{			0.04	}$ & $	1.63	\pm{	0.06			}$ & $	0.31	$&$	0.59	\pm{			0.09	}$ & $	6.16	\pm{			0.08	}$ & $	51.16	\pm{	4.02			}$ & $	87.91	\pm{	0.55			}$\\
15	&	50049-02-03-02	&	835.19034	&$	2.21	\pm{			0.06	}$ & $	1.49	\pm{	0.07			}$ & $	0.40	$&$	0.51	\pm{			0.11	}$ & $	4.33	\pm{	0.06			}$ & $	28.15	\pm{	2.96			}$ & $	50.67	\pm{	0.33			}$\\
16	&	50049-02-05-00	&	835.47605	&$	2.32	\pm{			0.06	}$ & $	1.65	\pm{	0.07			}$ & $	0.30	$&$	0.59	\pm{			0.14	}$ & $	6.58	\pm{			0.07	}$ & $	40.02	\pm{			4.54	}$ & $	90.68	\pm{	0.52			}$\\
17	&	50049-02-06-00	&	836.51501	&$	2.35	\pm{			0.08	}$ & $	1.66	\pm{	0.08			}$ & $	0.45	$&$	0.62	\pm{			0.18	}$ & $	6.73	\pm{			0.08	}$ & $	38.97	\pm{			5.37	}$ & $	92.11	\pm{	0.52			}$\\
18	&	50049-02-06-01	&	836.59062	&$	2.31	\pm{			0.08	}$ & $	1.65	\pm{	0.09			}$ & $	0.61	$&$	0.58	\pm{			0.19	}$ & $	6.47	\pm{			0.08	}$ & $	44.63	\pm{	6.76			}$ & $	89.74	\pm{	0.55			}$\\
19	&	50049-02-06-02	&	836.65903	&$	2.25	\pm{			0.09	}$ & $	1.58	\pm{	0.09			}$ & $	0.41	$&$	0.54	\pm{			0.19	}$ & $	5.53	\pm{			0.07	}$ & $	37.83	\pm{	6.14			}$ & $	79.84	\pm{	0.50			}$\\
20	&	50049-02-07-00	&	837.31072	&$	2.30	\pm{			0.07	}$ & $	1.64	\pm{	0.09			}$ & $	0.59	$&$	0.58	\pm{			0.16	}$ & $	6.34	\pm{			0.08	}$ & $	45.08	\pm{	5.90			}$ & $	86.81	\pm{	0.54			}$\\
21	&	50049-02-07-01	&	838.58253	&$	2.26	\pm{			0.08	}$ & $	1.56	\pm{	0.08			}$ & $	0.78	$&$	0.55	\pm{			0.16	}$ & $	5.21	\pm{			0.07	}$ & $	35.25	\pm{	4.99			}$ & $	75.20	\pm{	0.47			}$\\
22	&	50049-02-07-02	&	839.16547	&$	2.33	\pm{			0.07	}$ & $	1.66	\pm{	0.07			}$ & $	0.38	$&$	0.60	\pm{			0.16	}$ & $	6.59	\pm{			0.08	}$ & $	40.72	\pm{			5.23	}$ & $	88.67	\pm{	0.51			}$\\
23	&	50049-02-07-03	&	841.42330	&$	2.29	\pm{			0.07	}$ & $	1.52	\pm{	0.08			}$ & $	0.83	$&$	0.57	\pm{			0.14	}$ & $	4.67	\pm{			0.06	}$ & $	29.26	\pm{			3.45	}$ & $	63.35	\pm{	0.40			}$\\
24	&	50049-02-07-04	&	843.42348	&$	2.34	\pm{			0.08	}$ & $	1.65	\pm{	0.08			}$ & $	0.47	$&$	0.61	\pm{			0.19	}$ & $	6.52	\pm{			0.08	}$ & $	35.51	\pm{			5.09	}$ & $	79.77	\pm{	0.46			}$\\
25	&	50049-02-08-00	&	844.12055	&$	2.31	\pm{			0.05	}$ & $	1.62	\pm{	0.06			}$ & $	0.44	$&$	0.58	\pm{			0.13	}$ & $	6.06	\pm{			0.07	}$ & $	33.90	\pm{			3.53	}$ & $	74.70	\pm{	0.44			}$\\
26	&	50049-02-08-01	&	845.14572	&$	2.25	\pm{			0.06	}$ & $	1.53	\pm{	0.09			}$ & $	0.29	$&$	0.54	\pm{			0.11	}$ & $	4.80	\pm{			0.07	}$ & $	45.79	\pm{	4.64			}$ & $	64.78	\pm{	0.45			}$\\
27	&	50049-02-08-03	&	846.20867	&$	2.27	\pm{			0.07	}$ & $	1.56	\pm{	0.08			}$ & $	0.47	$&$	0.55	\pm{			0.15	}$ & $	5.20	\pm{			0.06	}$ & $	30.71	\pm{	3.89			}$ & $	65.45	\pm{	0.42			}$\\
28	&	50049-02-10-03	&	849.39406	&$	2.29	\pm{			0.11	}$ & $	1.57	\pm{	0.10			}$ & $	0.48	$&$	0.57	\pm{			0.24	}$ & $	5.37	\pm{			0.06	}$ & $	23.76	\pm{			4.69	}$ & $	60.09	\pm{	0.37			}$\\
29	&	50049-02-10-02	&	849.46316	&$	2.36	\pm{			0.11	}$ & $	1.65	\pm{	0.09			}$ & $	0.79	$&$	0.63	\pm{			0.26	}$ & $	6.47	\pm{			0.07	}$ & $	25.65	\pm{			4.95	}$ & $	68.85	\pm{	0.40			}$\\
30	&	50049-02-10-01	&	849.53465	&$	2.36	\pm{			0.11	}$ & $	1.63	\pm{	0.09			}$ & $	0.89	$&$	0.62	\pm{			0.24	}$ & $	6.26	\pm{			0.07	}$ & $	26.61	\pm{			4.62	}$ & $	67.68	\pm{	0.40			}$\\
31	&	50049-02-10-00	&	850.10229	&$	2.29	\pm{			0.05	}$ & $	1.61	\pm{	0.06			}$ & $	0.38	$&$	0.57	\pm{			0.12	}$ & $	5.92	\pm{			0.07	}$ & $	28.52	\pm{	2.95			}$ & $	63.79	\pm{	0.36			}$\\
32	&	50049-02-10-05	&	850.86400	&$	2.25	\pm{			0.06	}$ & $	1.56	\pm{	0.08			}$ & $	0.72	$&$	0.54	\pm{			0.14	}$ & $	5.18	\pm{			0.06	}$ & $	29.09	\pm{	3.53			}$ & $	56.65	\pm{	0.36			}$\\
33	&	50049-02-11-01	&	852.58772	&$	2.23	\pm{			0.11	}$ & $	1.54	\pm{	0.10			}$ & $	1.13	$&$	0.53	\pm{			0.23	}$ & $	5.01	\pm{			0.06	}$ & $	24.38	\pm{	4.74			}$ & $	53.92	\pm{	0.34			}$\\
34	&	50049-02-11-02	&	852.65313	&$	2.28	\pm{			0.16	}$ & $	1.60	\pm{	0.13			}$ & $	0.73	$&$	0.56	\pm{			0.34	}$ & $	5.70	\pm{			0.07	}$ & $	24.49	\pm{	5.96			}$ & $	57.21	\pm{	0.37			}$\\
35	&	50049-02-12-01	&	853.58329	&$	2.21	\pm{			0.09	}$ & $	1.49	\pm{	0.09			}$ & $	0.57	$&$	0.51	\pm{			0.15	}$ & $	4.35	\pm{			0.06	}$ & $	25.43	\pm{	3.74			}$ & $	48.45	\pm{	0.33			}$\\
36	&	50049-02-12-00	&	854.27232	&$	2.33	\pm{			0.06	}$ & $	1.54	\pm{	0.07			}$ & $	0.76	$&$	0.60	\pm{			0.13	}$ & $	4.90	\pm{			0.06	}$ & $	18.68	\pm{			2.09	}$ & $	45.47	\pm{	0.28			}$\\
37	&	50049-02-13-00	&	855.30361	&$	2.11	\pm{			0.04	}$ & $	1.18	\pm{	0.05			}$ & $	0.87	$&$	0.45	\pm{			0.04	}$ & $	1.40	\pm{			0.02	}$ & $	19.59	\pm{	0.99			}$ & $	25.38	\pm{	0.21			}$\\
38	&	50049-02-14-00	&	856.96964	&$	2.25	\pm{			0.05	}$ & $	1.42	\pm{	0.07			}$ & $	0.82	$&$	0.54	\pm{			0.10	}$ & $	3.54	\pm{			0.05	}$ & $	18.73	\pm{	1.78			}$ & $	34.56	\pm{	0.24			}$\\
39	&	50049-02-15-00	&	859.24090	&$	2.20	\pm{			0.05	}$ & $	1.37	\pm{	0.07			}$ & $	0.99	$&$	0.50	\pm{			0.08	}$ & $	2.97	\pm{			0.04	}$ & $	15.66	\pm{	1.44			}$ & $	25.45	\pm{	0.19			}$\\
40	&	50049-02-15-01	&	860.21521	&$	2.14	\pm{			0.10	}$ & $	1.41	\pm{			0.13	}$ & $	0.63	$&$	0.46	\pm{			0.16	}$ & $	3.35	\pm{			0.05	}$ & $	14.95	\pm{	2.78			}$ & $	25.42	\pm{	0.19			}$\\
41	&	50049-02-15-08	&	860.54742	&$	2.10	\pm{			0.10	}$ & $	1.18	\pm{			0.08	}$ & $	1.01	$&$	0.44	\pm{			0.10	}$ & $	1.43	\pm{			0.03	}$ & $	12.74	\pm{	1.58			}$ & $	15.82	\pm{	0.15			}$\\
42	&	50049-02-15-02	&	861.11940	&$	2.20	\pm{			0.06	}$ & $	1.34	\pm{			0.08	}$ & $	1.12	$&$	0.50	\pm{			0.08	}$ & $	2.71	\pm{			0.04	}$ & $	11.94	\pm{			0.89	}$ & $	19.12	\pm{	0.15			}$\\
43	&	50049-02-15-03	&	861.88606	&$	2.06	\pm{			0.04	}$ & $	1.12	\pm{			0.05	}$ & $	0.84	$&$	0.41	\pm{			0.04	}$ & $	1.05	\pm{			0.02	}$ & $	9.99	\pm{	0.53			}$ & $	11.14	\pm{	0.10			}$\\
44	&	50049-02-15-04	&	862.09305	&$	2.11	\pm{			0.05	}$ & $	1.13	\pm{			0.05	}$ & $	1.35	$&$	0.45	\pm{			0.05	}$ & $	1.10	\pm{			0.02	}$ & $	9.18	\pm{	0.57			}$ & $	10.45	\pm{	0.10			}$\\
45	&	50049-02-15-05	&	863.23997	&$	2.45	\pm{			0.08	}$ & $	1.28	\pm{			0.06	}$ & $	1.72	$&$	0.70	\pm{			0.09	}$ & $	2.13	\pm{			0.03	}$ & $	5.78	\pm{	0.39			}$ & $	9.03	\pm{	0.08			}$\\
46	&	50049-02-15-06	&	864.20819	&$	2.12	\pm{			0.11	}$ & $	1.16	\pm{			0.08	}$ & $	0.83	$&$	0.45	\pm{			0.10	}$ & $	1.27	\pm{			0.03	}$ & $	6.07	\pm{	0.67			}$ & $	7.24	\pm{	0.07			}$\\
47	&	50049-02-15-07	&	864.27747	&$	2.24	\pm{			0.15	}$ & $	1.17	\pm{			0.08	}$ & $	0.85	$&$	0.54	\pm{			0.14	}$ & $	1.31	\pm{			0.03	}$ & $	5.46	\pm{	0.70			}$ & $	7.19	\pm{	0.08			}$\\
\hline
48	&	50049-03-01-00	&	865.27217	&$	4.43	\pm{			0.34	}$ & $	1.86	\pm{	0.07			}$ & $	1.48	$&$	3.10	\pm{			0.61	}$ & $	10.09	\pm{			0.17	}$ & $	3.99	\pm{			0.17	}$ & $	5.83	\pm{	0.05			}$\\
49	&	50049-03-02-01	&	866.92250	&$	4.73	\pm{			0.46	}$ & $	1.74	\pm{	0.11			}$ & $	0.99	$&$	3.53	\pm{	0.63			}$ & $	7.87	\pm{			0.25	}$ & $	3.64	\pm{	0.16			}$ & $	2.63	\pm{	0.04			}$\\
50	&	50049-03-02-00	&	866.97790	&$	6.05	\pm{	0.89			}$ & $	1.67	\pm{	0.17			}$ & $	0.96	$&$	5.55	\pm{	4.06			}$ & $	6.88	\pm{			0.15	}$ & $	3.55	\pm{	1.55			}$ & $	2.45	\pm{	0.03			}$\\
51	&	50049-03-03-00	&	868.23607	&$	6.56	\pm{	1.37			}$ & $	1.44	\pm{			0.16	}$ & $	0.74	$&$	6.31	\pm{	4.82			}$ & $	3.75	\pm{			0.14	}$ & $	2.17	\pm{	1.49			}$ & $	1.12	\pm{	0.02			}$\\
52	&	50049-03-04-00	&	869.47435	&$	6.88	\pm{	1.28			}$ & $	1.32	\pm{			0.20	}$ & $	0.86	$&$	6.78	\pm{	4.38			}$ & $	2.49	\pm{			0.11	}$ & $	0.95	\pm{	0.47			}$ & $	0.38	\pm{	0.01			}$\\
\hline
\end{tabular}
\end{center}
\footnotesize{ 
\begin{flushleft} $^{a}$ Degree of freedom (d.o.f) is 48 for all observation.\\
$^{b}$ Hardness parameters are obtained using the flux ratio of two different energy ranges; $F(10-30~{\rm keV})/F(3-10~{\rm keV})$.
The parameters for diskbb component are multiplied by $10^2$ because of low values\\
$^{c}$ Unabsorbed fluxes of the blackbody components are in units of $10^{-10}~{\rm erg~ s^{-1}~cm^{-2}}$. \\
$^d$ Unabsorbed fluxes of the disk blackbody components are in units of $10^{-10}~{\rm erg~s^{-1}~cm^{-2}}$.
\end{flushleft} }
\label{bbdiskbb.2000}
\end{table*}

\begin{table*}
\caption{Same as Table \ref{bbdiskbb.2000} but for the 2011 outburst.}
\begin{center}
\tiny
\begin{tabular}{cccccccccccccccccccccc}
\hline
\hline
Obs \#	&	ObsID	&	MJD$-55000$	&	$kT_{bbody}$					&$	 kT_{diskbb}					$&	$\chi^2/d.o.f.^{a}$	&	Hardness$^{b}$					&	Hardness$^{b}$					&	Flux$^{c}_{bbody}$					&	Flux$^{d}_{diskbb}$					\\
	&		&	(days)	&	(keV)					&	(keV)					&		&	bbody					&	diskbb					&						&						\\ \hline
1	&	96440-01-01-00	&	849.3771	&$	8.04	\pm{	1.71			}$ & $	1.63	\pm{			0.07	}$ & $	1.09	$&$	8.39	\pm{			0.32	}$ & $	6.19	\pm{			0.12	}$ & $	3.74	\pm{			0.07	}$ & $	2.76	\pm{			0.03	}$\\
2	&	96440-01-01-01	&	852.4442	&$	12.11	\pm{	4.58			}$ & $	2.06	\pm{			0.05	}$ & $	0.98	$&$	12.77	\pm{			0.42	}$ & $	14.08	\pm{			0.19	}$ & $	6.83	\pm{			0.11	}$ & $	9.78	\pm{			0.06	}$\\
3	&	96440-01-02-02	&	855.5727	&$	5.33	\pm{			0.13	}$ & $	2.45	\pm{			0.05	}$ & $	1.11	$&$	4.45	\pm{			0.06	}$ & $	23.44	\pm{			0.26	}$ & $	31.11	\pm{			0.19	}$ & $	30.63	\pm{			0.17	}$\\
\hline
4	&	96440-01-02-00	&	858.1840	&$	2.24	\pm{			0.05	}$ & $	1.55	\pm{			0.06	}$ & $	0.60	$&$	0.53	\pm{			0.01	}$ & $	5.10	\pm{			0.06	}$ & $	43.52	\pm{			0.25	}$ & $	91.63	\pm{			0.54	}$\\
5	&	96440-01-02-03	&	859.6329	&$	2.32	\pm{			0.05	}$ & $	1.57	\pm{			0.08	}$ & $	0.54	$&$	0.59	\pm{			0.01	}$ & $	5.36	\pm{			0.07	}$ & $	74.03	\pm{			0.41	}$ & $	98.06	\pm{			0.69	}$\\
6	&	96440-01-02-01	&	861.2549	&$	2.50	\pm{			0.03	}$ & $	1.69	\pm{			0.01	}$ & $	1.10	$&$	0.74	\pm{			0.01	}$ & $	3.97	\pm{			0.20	}$ & $	26.10	\pm{			0.51	}$ & $	78.60	\pm{			0.34	}$\\
7	&	96440-01-03-02	&	862.4965	&$	2.24	\pm{			0.06	}$ & $	1.55	\pm{			0.07	}$ & $	0.54	$&$	0.53	\pm{			0.01	}$ & $	5.12	\pm{			0.06	}$ & $	50.02	\pm{			0.32	}$ & $	102.58	\pm{			0.63	}$\\
8	&	96440-01-03-05	&	863.3270	&$	2.25	\pm{			0.07	}$ & $	1.57	\pm{			0.07	}$ & $	0.71	$&$	0.54	\pm{			0.01	}$ & $	5.32	\pm{			0.07	}$ & $	52.41	\pm{			0.35	}$ & $	108.54	\pm{			0.67	}$\\
9	&	96440-01-03-00	&	864.2559	&$	2.24	\pm{			0.06	}$ & $	1.57	\pm{			0.07	}$ & $	0.61	$&$	0.53	\pm{			0.01	}$ & $	5.41	\pm{			0.07	}$ & $	54.97	\pm{			0.32	}$ & $	107.14	\pm{			0.65	}$\\
10	&	96440-01-03-03	&	865.4324	&$	2.34	\pm{			0.08	}$ & $	1.66	\pm{			0.06	}$ & $	0.42	$&$	0.61	\pm{			0.01	}$ & $	6.66	\pm{			0.07	}$ & $	47.21	\pm{			0.33	}$ & $	119.18	\pm{			0.65	}$\\
11	&	96440-01-03-06	&	866.2909	&$	2.30	\pm{			0.08	}$ & $	1.63	\pm{			0.08	}$ & $	0.52	$&$	0.58	\pm{			0.01	}$ & $	6.14	\pm{			0.07	}$ & $	56.66	\pm{			0.40	}$ & $	118.43	\pm{			0.73	}$\\
12	&	96440-01-03-01	&	867.2569	&$	2.30	\pm{			0.06	}$ & $	1.64	\pm{			0.06	}$ & $	0.38	$&$	0.58	\pm{			0.01	}$ & $	6.37	\pm{			0.07	}$ & $	57.38	\pm{			0.32	}$ & $	118.24	\pm{			0.68	}$\\
13	&	96440-01-03-04	&	867.9826	&$	2.32	\pm{			0.07	}$ & $	1.68	\pm{			0.07	}$ & $	0.37	$&$	0.60	\pm{			0.01	}$ & $	7.02	\pm{			0.08	}$ & $	56.88	\pm{			0.37	}$ & $	125.92	\pm{			0.74	}$\\
14	&	96440-01-04-03	&	869.2781	&$	2.26	\pm{			0.07	}$ & $	1.58	\pm{			0.07	}$ & $	0.53	$&$	0.55	\pm{			0.01	}$ & $	5.44	\pm{			0.07	}$ & $	51.59	\pm{			0.35	}$ & $	106.77	\pm{			0.65	}$\\
15	&	96440-01-04-04	&	871.2958	&$	2.35	\pm{			0.08	}$ & $	1.69	\pm{			0.07	}$ & $	0.55	$&$	0.61	\pm{			0.01	}$ & $	7.16	\pm{			0.08	}$ & $	48.77	\pm{			0.33	}$ & $	118.72	\pm{			0.66	}$\\
16	&	96440-01-04-01	&	872.0858	&$	2.22	\pm{			0.06	}$ & $	1.54	\pm{			0.07	}$ & $	0.41	$&$	0.52	\pm{			0.01	}$ & $	4.90	\pm{			0.06	}$ & $	46.96	\pm{			0.29	}$ & $	93.05	\pm{			0.57	}$\\
17	&	96440-01-04-05	&	873.0034	&$	2.27	\pm{			0.04	}$ & $	1.53	\pm{			0.07	}$ & $	0.49	$&$	0.56	\pm{			0.01	}$ & $	4.78	\pm{			0.07	}$ & $	80.36	\pm{			0.41	}$ & $	100.79	\pm{			0.74	}$\\
18	&	96440-01-04-06	&	875.6732	&$	2.31	\pm{			0.08	}$ & $	1.65	\pm{			0.08	}$ & $	0.56	$&$	0.59	\pm{			0.01	}$ & $	6.55	\pm{			0.07	}$ & $	46.82	\pm{			0.32	}$ & $	110.27	\pm{			0.61	}$\\
19	&	96440-01-05-00	&	876.8402	&$	2.27	\pm{			0.04	}$ & $	1.55	\pm{			0.08	}$ & $	0.49	$&$	0.55	\pm{			0.01	}$ & $	5.10	\pm{			0.07	}$ & $	72.06	\pm{			0.34	}$ & $	91.93	\pm{			0.64	}$\\
20	&	96440-01-05-01	&	877.4295	&$	2.32	\pm{			0.07	}$ & $	1.63	\pm{			0.06	}$ & $	0.59	$&$	0.60	\pm{			0.01	}$ & $	6.21	\pm{			0.07	}$ & $	44.29	\pm{			0.29	}$ & $	101.79	\pm{			0.58	}$\\
21	&	96440-01-05-02	&	877.8880	&$	2.24	\pm{			0.05	}$ & $	1.54	\pm{			0.07	}$ & $	0.57	$&$	0.53	\pm{			0.01	}$ & $	5.00	\pm{			0.06	}$ & $	53.36	\pm{			0.30	}$ & $	91.42	\pm{			0.59	}$\\
22	&	96440-01-05-03	&	879.5833	&$	2.24	\pm{			0.07	}$ & $	1.56	\pm{			0.07	}$ & $	0.54	$&$	0.54	\pm{			0.01	}$ & $	5.21	\pm{			0.06	}$ & $	38.17	\pm{			0.25	}$ & $	82.47	\pm{			0.49	}$\\
23	&	96440-01-05-04	&	880.6295	&$	2.27	\pm{			0.08	}$ & $	1.59	\pm{			0.07	}$ & $	0.70	$&$	0.55	\pm{			0.01	}$ & $	5.67	\pm{			0.07	}$ & $	38.00	\pm{			0.26	}$ & $	86.76	\pm{			0.48	}$\\
24	&	96440-01-05-05	&	882.6413	&$	2.23	\pm{			0.08	}$ & $	1.56	\pm{			0.08	}$ & $	0.57	$&$	0.53	\pm{			0.01	}$ & $	5.20	\pm{			0.07	}$ & $	44.76	\pm{			0.30	}$ & $	83.84	\pm{			0.53	}$\\
25	&	96440-01-06-00	&	883.5596	&$	2.27	\pm{			0.07	}$ & $	1.58	\pm{			0.07	}$ & $	0.59	$&$	0.56	\pm{			0.01	}$ & $	5.56	\pm{			0.07	}$ & $	44.31	\pm{			0.28	}$ & $	84.61	\pm{			0.51	}$\\
26	&	96440-01-06-01	&	884.6704	&$	2.18	\pm{			0.05	}$ & $	1.45	\pm{			0.07	}$ & $	0.88	$&$	0.49	\pm{			0.01	}$ & $	3.79	\pm{			0.05	}$ & $	35.19	\pm{			0.21	}$ & $	61.04	\pm{			0.40	}$\\
27	&	96440-01-06-02	&	885.8371	&$	2.19	\pm{			0.05	}$ & $	1.44	\pm{			0.07	}$ & $	0.60	$&$	0.50	\pm{			0.01	}$ & $	3.75	\pm{			0.05	}$ & $	36.20	\pm{			0.20	}$ & $	59.89	\pm{			0.40	}$\\
28	&	96440-01-06-03	&	886.8228	&$	2.22	\pm{			0.05	}$ & $	1.50	\pm{			0.08	}$ & $	0.73	$&$	0.52	\pm{			0.01	}$ & $	4.38	\pm{			0.06	}$ & $	47.34	\pm{			0.24	}$ & $	65.07	\pm{			0.42	}$\\
29	&	96440-01-06-04	&	888.2533	&$	2.20	\pm{			0.06	}$ & $	1.42	\pm{			0.07	}$ & $	0.65	$&$	0.50	\pm{			0.01	}$ & $	3.48	\pm{			0.05	}$ & $	29.03	\pm{			0.17	}$ & $	52.16	\pm{			0.35	}$\\
30	&	96440-01-07-00	&	890.0729	&$	2.22	\pm{			0.07	}$ & $	1.44	\pm{			0.08	}$ & $	0.81	$&$	0.52	\pm{			0.01	}$ & $	3.72	\pm{			0.05	}$ & $	26.97	\pm{			0.17	}$ & $	48.77	\pm{			0.32	}$\\
31	&	96440-01-07-01	&	891.1884	&$	2.26	\pm{			0.08	}$ & $	1.54	\pm{			0.08	}$ & $	0.85	$&$	0.55	\pm{			0.01	}$ & $	4.94	\pm{			0.06	}$ & $	25.15	\pm{			0.17	}$ & $	54.51	\pm{			0.32	}$\\
32	&	96440-01-07-02	&	892.2633	&$	2.25	\pm{			0.09	}$ & $	1.54	\pm{			0.08	}$ & $	0.41	$&$	0.54	\pm{			0.01	}$ & $	4.99	\pm{			0.06	}$ & $	24.98	\pm{			0.17	}$ & $	53.47	\pm{			0.31	}$\\
33	&	96440-01-07-03	&	893.4686	&$	2.12	\pm{			0.06	}$ & $	1.40	\pm{			0.08	}$ & $	0.67	$&$	0.45	\pm{			0.01	}$ & $	3.28	\pm{			0.05	}$ & $	29.27	\pm{			0.18	}$ & $	43.78	\pm{			0.31	}$\\
34	&	96440-01-07-04	&	894.3870	&$	2.15	\pm{			0.05	}$ & $	1.33	\pm{			0.07	}$ & $	0.84	$&$	0.47	\pm{			0.01	}$ & $	2.54	\pm{			0.04	}$ & $	21.01	\pm{			0.13	}$ & $	32.87	\pm{			0.23	}$\\
35	&	96440-01-07-05	&	895.2340	&$	2.17	\pm{			0.06	}$ & $	1.37	\pm{			0.08	}$ & $	0.72	$&$	0.48	\pm{			0.01	}$ & $	2.95	\pm{			0.04	}$ & $	21.07	\pm{			0.13	}$ & $	34.92	\pm{			0.23	}$\\
36	&	96440-01-07-06	&	896.1426	&$	2.07	\pm{			0.04	}$ & $	1.29	\pm{			0.08	}$ & $	0.80	$&$	0.41	\pm{			0.01	}$ & $	2.18	\pm{			0.04	}$ & $	28.16	\pm{			0.15	}$ & $	30.50	\pm{			0.25	}$\\
37	&	96440-01-08-00	&	898.1645	&$	2.06	\pm{			0.04	}$ & $	1.10	\pm{			0.04	}$ & $	1.35	$&$	0.41	\pm{			0.01	}$ & $	0.97	\pm{			0.02	}$ & $	14.75	\pm{			0.09	}$ & $	17.27	\pm{			0.15	}$\\
38	&	96440-01-08-02	&	899.0077	&$	2.12	\pm{			0.06	}$ & $	1.22	\pm{			0.07	}$ & $	1.54	$&$	0.45	\pm{			0.01	}$ & $	1.65	\pm{			0.03	}$ & $	15.29	\pm{			0.11	}$ & $	19.47	\pm{			0.16	}$\\
39	&	96440-01-08-03	&	900.3845	&$	2.12	\pm{			0.08	}$ & $	1.20	\pm{			0.06	}$ & $	1.31	$&$	0.45	\pm{			0.01	}$ & $	1.53	\pm{			0.03	}$ & $	11.34	\pm{			0.08	}$ & $	15.41	\pm{			0.13	}$\\
40	&	96440-01-08-04	&	901.2278	&$	2.10	\pm{			0.06	}$ & $	1.23	\pm{			0.07	}$ & $	0.92	$&$	0.44	\pm{			0.01	}$ & $	1.75	\pm{			0.03	}$ & $	12.26	\pm{			0.09	}$ & $	15.87	\pm{			0.12	}$\\
41	&	96440-01-08-05	&	902.0756	&$	2.02	\pm{			0.05	}$ & $	1.14	\pm{			0.06	}$ & $	0.78	$&$	0.39	\pm{			0.01	}$ & $	1.15	\pm{			0.02	}$ & $	11.22	\pm{			0.08	}$ & $	12.26	\pm{			0.11	}$\\
42	&	96440-01-08-06	&	903.1209	&$	1.94	\pm{			0.05	}$ & $	1.03	\pm{			0.05	}$ & $	0.86	$&$	0.34	\pm{			0.01	}$ & $	0.63	\pm{			0.01	}$ & $	7.86	\pm{			0.06	}$ & $	7.10	\pm{			0.08	}$\\
43	&	96440-01-09-00	&	904.0978	&$	2.29	\pm{			0.10	}$ & $	1.17	\pm{			0.06	}$ & $	1.56	$&$	0.57	\pm{			0.01	}$ & $	1.34	\pm{			0.03	}$ & $	4.44	\pm{			0.05	}$ & $	5.74	\pm{			0.05	}$\\
44	&	96440-01-09-08	&	904.3096	&$	2.31	\pm{			0.10	}$ & $	1.21	\pm{			0.06	}$ & $	1.35	$&$	0.58	\pm{			0.01	}$ & $	1.63	\pm{			0.03	}$ & $	3.95	\pm{			0.04	}$ & $	5.86	\pm{			0.05	}$\\
\hline
45	&	96440-01-09-02	&	906.0508	&$	9.22	\pm{	3.94			}$ & $	1.81	\pm{			0.06	}$ & $	0.93	$&$	9.85	\pm{			0.72	}$ & $	9.19	\pm{			0.17	}$ & $	2.15	\pm{			0.08	}$ & $	3.47	\pm{			0.03	}$\\
46	&	96440-01-09-03	&	907.3148	&$	15.80	\pm{	7.33			}$ & $	1.66	\pm{			0.05	}$ & $	1.17	$&$	15.40	\pm{			0.96	}$ & $	6.67	\pm{			0.12	}$ & $	1.56	\pm{			0.05	}$ & $	1.95	\pm{			0.02	}$\\
47	&	96440-01-09-10	&	908.0061	&$	7.94	\pm{	2.18			}$ & $	1.51	\pm{			0.11	}$ & $	1.03	$&$	8.26	\pm{			0.52	}$ & $	4.63	\pm{			0.13	}$ & $	2.05	\pm{			0.06	}$ & $	1.40	\pm{			0.02	}$\\
48	&	96440-01-09-04	&	908.2906	&$	9.27	\pm{	2.57			}$ & $	1.49	\pm{			0.06	}$ & $	1.02	$&$	9.92	\pm{			0.49	}$ & $	4.37	\pm{			0.09	}$ & $	1.69	\pm{			0.04	}$ & $	1.22	\pm{			0.01	}$\\
49	&	96440-01-09-05	&	909.0731	&$	6.65	\pm{			1.80	}$ & $	1.30	\pm{			0.09	}$ & $	1.07	$&$	6.44	\pm{			0.27	}$ & $	2.27	\pm{			0.06	}$ & $	1.56	\pm{			0.03	}$ & $	0.75	\pm{			0.01	}$\\
50	&	96440-01-09-11	&	909.2486	&$	7.83	\pm{	2.14			}$ & $	1.35	\pm{			0.12	}$ & $	0.84	$&$	8.11	\pm{			0.59	}$ & $	2.78	\pm{			0.10	}$ & $	1.41	\pm{			0.05	}$ & $	0.72	\pm{			0.01	}$\\
51	&	96440-01-09-12	&	910.9905	&$	3.72	\pm{	3.72			}$ & $	1.11	\pm{			0.91	}$ & $	0.61	$&$	2.09	\pm{			0.72	}$ & $	0.76	\pm{			0.27	}$ & $	0.45	\pm{			0.07	}$ & $	0.16	\pm{			0.03	}$\\
\hline
\end{tabular}
\end{center}
\footnotesize{ 
\begin{flushleft} $^{a}$ Degree of freedom (d.o.f) is 49 for all observation.\\
$^{b}$ Hardness parameters are obtained using the flux ratio of two different energy ranges; $F(10-30~{\rm keV})/F(3-10~{\rm keV})$.
The parameters for disk blackbody component are multiplied by $10^2$ because of low values\\
$^{c}$ Unabsorbed fluxes of the blackbody components are in units of $10^{-10}~{\rm erg~ s^{-1}~cm^{-2}}$. \\
$^d$ Unabsorbed fluxes of the disk blackbody components are in units of $10^{-10}~{\rm erg~s^{-1}~cm^{-2}}$.
\end{flushleft} }
\label{bbdiskbb.2011}
\end{table*}

\begin{table*}
\caption{Best fit parameters of \textit{blackbody + comptonization + disk blackbody + Gauss} model for the 2000 outburst.
The horizontal lines in the table indicate the state transitions according to hardness parameter of the blackbody component.}
\begin{center}
\tiny
\begin{tabular}{cccccccccccccc}
\hline
\hline
Obs \#	&	$kT_{bbody}$					&$	\tau					$&	$\chi^2/d.o.f.^a$	&	Hardness$^{b}$					&	Flux$^{c}_{bbody}$					&	\\
	&	(keV)					&						&		&	bbody					&						&	\\ \hline
1	&$	2.36	\pm{			0.30	}$ & $	2.72	\pm{			0.67	}$ & $	0.95	$&$	2.77	\pm{			0.10	}$ & $	1.84	\pm{			0.02	}	$\\
2	&$	2.80	\pm{			0.21	}$ & $	2.49	\pm{			0.30	}$ & $	1.13	$&$	3.30	\pm{			0.09	}$ & $	3.37	\pm{			0.03	}	$\\
3	&$	2.77	\pm{			0.21	}$ & $	2.76	\pm{			0.26	}$ & $	1.03	$&$	3.49	\pm{			0.08	}$ & $	5.10	\pm{			0.04	}	$\\
4	&$	3.61	\pm{	0.36			}$ & $	2.38	\pm{	0.60			}$ & $	0.68	$&$	4.73	\pm{	0.07			}$ & $	16.74	\pm{		0.07		}	$\\
5	&$	3.44	\pm{	0.70			}$ & $	2.44	\pm{	0.16			}$ & $	0.69	$&$	4.31	\pm{	0.06			}$ & $	24.74	\pm{		0.10		}	$\\
\hline
6	&$	2.17	\pm{		0.13		}$ & $	0.12	\pm{	0.04			}$ & $	0.63	$&$	0.59	\pm{	0.01			}$ & $	30.91	\pm{		0.17		}	$\\
7	&$	2.34	\pm{	0.05			}$ & $	0.01	\pm{	0.01			}$ & $	0.62	$&$	0.65	\pm{	0.01			}$ & $	34.82	\pm{		0.20		}	$\\
8	&$	2.24	\pm{	0.03			}$ & $	0.10	\pm{	0.02			}$ & $	0.32	$&$	0.56	\pm{	0.01			}$ & $	48.09	\pm{		0.21		}	$\\
9	&$	2.22	\pm{	0.05			}$ & $	0.01	\pm{	0.01			}$ & $	0.68	$&$	0.60	\pm{	0.01			}$ & $	30.65	\pm{		0.17		}	$\\
10	&$	2.27	\pm{	0.03			}$ & $	0.01	\pm{	0.01			}$ & $	0.51	$&$	0.60	\pm{	0.01			}$ & $	44.13	\pm{		0.20		}	$\\
16	&$	2.25	\pm{			0.03	}$ & $	0.13	\pm{	0.09			}$ & $	0.52	$&$	0.58	\pm{			0.01	}$ & $	40.97	\pm{			0.20	}	$\\
31	&$	2.25	\pm{	0.04			}$ & $	0.11	\pm{	0.10			}$ & $	0.49	$&$	0.56	\pm{			0.01	}$ & $	28.96	\pm{			0.14	}	$\\
39	&$	2.00	\pm{			0.03	}$ & $	0.06	\pm{			0.02	}$ & $	0.93	$&$	0.48	\pm{			0.01	}$ & $	16.36	\pm{			0.07	}	$\\
41	&$	1.91	\pm{	0.03			}$ & $	0.03	\pm{	0.02			}$ & $	1.30	$&$	0.40	\pm{	0.01			}$ & $	13.82	\pm{		0.07		}	$\\
44	&$	1.70	\pm{	0.06			}$ & $	0.45	\pm{	0.36			}$ & $	1.74	$&$	0.34	\pm{	0.01			}$ & $	12.04	\pm{		0.05		}	$\\
45	&$	1.80	\pm{	0.08			}$ & $	0.48	\pm{	0.13			}$ & $	1.99	$&$	0.49	\pm{	0.01			}$ & $	7.85	\pm{		0.03		}	$\\
46	&$	1.95	\pm{	0.04			}$ & $	0.01	\pm{	0.01			}$ & $	1.63	$&$	0.40	\pm{	0.01			}$ & $	6.74	\pm{		0.04		}	$\\
47	&$	2.01	\pm{	0.15			}$ & $	0.11	\pm{	0.07			}$ & $	0.80	$&$	0.53	\pm{	0.01			}$ & $	5.65	\pm{		0.04		}	$\\
\hline
48	&$	3.48	\pm{			0.45	}$ & $	1.63	\pm{	0.51			}$ & $	1.40	$&$	3.27	\pm{			0.13	}$ & $	4.06	\pm{			0.05	}	$\\
49	&$	3.75	\pm{	0.65			}$ & $	2.30	\pm{	0.88			}$ & $	0.93	$&$	3.71	\pm{			0.19	}$ & $	3.71	\pm{			0.07	}	$\\
50	&$	2.82	\pm{			0.40	}$ & $	2.83	\pm{	0.22			}$ & $	0.91	$&$	3.64	\pm{			0.11	}$ & $	4.00	\pm{			0.04	}	$\\
51	&$	2.72	\pm{			0.48	}$ & $	2.84	\pm{			0.85	}$ & $	0.67	$&$	3.47	\pm{			0.15	}$ & $	2.49	\pm{			0.04	}	$\\
52	&$	2.55	\pm{			0.56	}$ & $	2.98	\pm{	0.20			}$ & $	0.78	$&$	3.29	\pm{			0.15	}$ & $	1.07	\pm{			0.02	}	$\\
\hline
\end{tabular}
\end{center}
\footnotesize{ 
\begin{flushleft}  $^{a}$ Degree of freedom (d.o.f) is 49 for all observation.\\
$^{b}$ Hardness parameters are obtained using the flux ratio of two different energy ranges; $F(10-30~{\rm keV})/F(3-10~{\rm keV})$.\\
$^{c}$ Unabsorbed fluxes of the blackbody components are in units of $10^{-10}~{\rm erg~ s^{-1}~cm^{-2}}$.
\end{flushleft} }
\label{comptt.2000}
\end{table*}

\begin{table*}
\caption{Same as \autoref{comptt.2000} but for the 2011 outburst.}
\begin{center}
\tiny
\begin{tabular}{cccccccccccccccc}
\hline
\hline
Obs \#	&	$kT_{bbody}$					&$	\tau					$&	$\chi^2/d.o.f.^{a}$	&	Hardness$^{b}$					&	Flux$^{c}_{bbody}$				&	\\
	&	(keV)					&						&		&	bbody					&					&	\\ \hline
1	&$	2.83	\pm{			0.31	}$ & $	2.57	\pm{			0.45	}$ & $	1.18	$&$	3.36	\pm{			0.07	}$ & $	5.27	\pm{			0.04	}$\\
2	&$	3.82	\pm{		0.40		}$ & $	2.16	\pm{		0.77		}$ & $	1.08	$&$	4.94	\pm{		0.13		}$ & $	13.76	\pm{		0.10		}$\\
3	&$	2.98	\pm{		0.58		}$ & $	1.91	\pm{		0.85		}$ & $	0.96	$&$	3.07	\pm{		0.05		}$ & $	37.48	\pm{		0.19		}$\\
\hline
4	&$	2.21	\pm{	0.04			}$ & $	0.01	\pm{	0.01			}$ & $	0.67	$&$	0.54	\pm{	0.01			}$ & $	43.37	\pm{	0.21			}$\\
5	&$	2.29	\pm{	0.03			}$ & $	0.01	\pm{	0.01			}$ & $	0.66	$&$	0.60	\pm{	0.01			}$ & $	73.84	\pm{		0.30		}$\\
6	&$	2.47	\pm{	0.01			}$ & $	0.01	\pm{	0.01			}$ & $	1.79	$&$	0.71	\pm{	0.01			}$ & $	26.81	\pm{	0.17			}$\\
7	&$	2.17	\pm{	0.04			}$ & $	0.01	\pm{	0.01			}$ & $	0.72	$&$	0.54	\pm{	0.01			}$ & $	49.46	\pm{		0.26		}$\\
8	&$	2.18	\pm{	0.04			}$ & $	0.07	\pm{	0.05			}$ & $	0.92	$&$	0.53	\pm{		0.01		}$ & $	53.56	\pm{		0.28		}$\\
9	&$	2.19	\pm{			0.03	}$ & $	0.09	\pm{	0.08			}$ & $	0.74	$&$	0.52	\pm{			0.01	}$ & $	55.90	\pm{			0.27	}$\\
10	&$	2.28	\pm{	0.05			}$ & $	0.21	\pm{	0.21			}$ & $	0.52	$&$	0.59	\pm{	0.01			}$ & $	48.14	\pm{		0.27		}$\\
12	&$	2.25	\pm{	0.03			}$ & $	0.01	\pm{	0.01			}$ & $	0.60	$&$	0.59	\pm{	0.01			}$ & $	56.81	\pm{		0.28		}$\\
19	&$	2.21	\pm{			0.03	}$ & $	0.06	\pm{	0.04			}$ & $	0.86	$&$	0.54	\pm{			0.01	}$ & $	73.73	\pm{			0.28	}$\\
30	&$	2.06	\pm{			0.05	}$ & $	0.07	\pm{	0.03			}$ & $	0.75	$&$	0.51	\pm{			0.01	}$ & $	27.71	\pm{			0.14	}$\\
40	&$	1.83	\pm{	0.10			}$ & $	0.34	\pm{	0.23			}$ & $	0.95	$&$	0.37	\pm{	0.00			}$ & $	14.42	\pm{		0.07		}$\\
41	&$	1.84	\pm{	0.02			}$ & $	0.04	\pm{	0.02			}$ & $	1.09	$&$	0.37	\pm{	0.01			}$ & $	11.89	\pm{	0.06			}$\\
42	&$	1.67	\pm{		0.05		}$ & $	0.10	\pm{		0.01		}$ & $	0.91	$&$	0.30	\pm{	0.00			}$ & $	8.92	\pm{		0.04		}$\\
43	&$	1.80	\pm{	0.01			}$ & $	0.40	\pm{	0.03			}$ & $	1.35	$&$	0.45	\pm{	0.01			}$ & $	5.52	\pm{	0.03			}$\\
44	&$	1.64	\pm{	0.04			}$ & $	0.34	\pm{	0.03			}$ & $	1.76	$&$	0.41	\pm{	0.01			}$ & $	5.50	\pm{	0.03			}$\\
\hline
45	&$	3.60	\pm{			0.60	}$ & $	2.66	\pm{	0.60			}$ & $	0.91	$&$	3.67	\pm{			0.12	}$ & $	3.64	\pm{			0.04	}$\\
46	&$	3.17	\pm{			0.47	}$ & $	3.15	\pm{	0.48			}$ & $	1.32	$&$	3.55	\pm{			0.04	}$ & $	3.06	\pm{			0.02	}$\\
47	&$	2.51	\pm{	0.24			}$ & $	3.00	\pm{	3.00			}$ & $	1.07	$&$	2.89	\pm{	0.09			}$ & $	2.73	\pm{		0.03		}$\\
48	&$	2.70	\pm{			0.30	}$ & $	2.68	\pm{	0.17			}$ & $	1.17	$&$	3.28	\pm{			0.08	}$ & $	2.40	\pm{			0.02	}$\\
49	&$	2.14	\pm{		0.16		}$ & $	2.99	\pm{	0.05			}$ & $	1.24	$&$	2.54	\pm{	0.06			}$ & $	1.91	\pm{		0.02		}$\\
50	&$	2.39	\pm{		0.25		}$ & $	3.01	\pm{	0.08			}$ & $	0.84	$&$	2.75	\pm{	0.09			}$ & $	1.77	\pm{		0.02		}$\\
51	&$	3.24	\pm{	1.59			}$ & $	2.74	\pm{	2.61			}$ & $	0.56	$&$	1.82	\pm{	0.44			}$ & $	0.47	\pm{		0.06		}$\\
\hline
\end{tabular}
\end{center}
\footnotesize{ 
\begin{flushleft}  $^{a}$ Degree of freedom (d.o.f) is 50 for all observation.\\
$^{b}$ Hardness parameters are obtained using the flux ratio of two different energy ranges; $F(10-30~{\rm keV})/F(3-10~{\rm keV})$.\\
$^{c}$ Unabsorbed fluxes of the blackbody components are in units of $10^{-10}~{\rm erg~ s^{-1}~cm^{-2}}$.
\end{flushleft} }
\label{comptt.2011}
\end{table*}

\end{document}